\documentclass[12pt,reprint,aps,nofootinbib,prd,twocolumn]{revtex4-2}

\usepackage[english]{babel}
\usepackage[utf8]{inputenc}
\usepackage{amsmath}
\usepackage{mathbbol}
\usepackage{amssymb}
\usepackage{tabularx}
\usepackage[normalem]{ulem}
\usepackage{bbold}
\usepackage{graphicx,amsfonts}
\usepackage{epsfig}
\usepackage[colorlinks=true,
linkcolor=blue,
urlcolor=blue,
citecolor=blue]{hyperref}
\usepackage{bm,color}
\usepackage[dvipsnames]{xcolor}
\usepackage{mathrsfs}
\usepackage{enumerate}
\usepackage{amsthm}
\usepackage{bbm}
\usepackage{comment}
\usepackage{physics}
\usepackage{url}

\begin{document}

\title{Spectral (in)stability of quasinormal modes and strong cosmic censorship}

\author{Aubin Courty$^{1,2,3}$, Kyriakos Destounis$^{1,3}$ and Paolo Pani$^{1,3}$} 
\affiliation{$^1$Dipartimento di Fisica, Sapienza Università di Roma, Piazzale Aldo Moro 5, 00185, Roma, Italy}
\affiliation{$^2$Laboratoire de Physique, École normale supérieure de Lyon - UMR 5672, Lyon, France}
\affiliation{$^3$INFN, Sezione di Roma, Piazzale Aldo Moro 2, 00185, Roma, Italy} 

\begin{abstract}
Recent studies have shown that quasinormal modes suffer from spectral instabilities, a frailty of black holes that leads to disproportional migration of their spectra in the complex plane when black-hole effective potentials are modified by minuscule perturbations. Similar results have been found with the mathematical notion of the pseudospectrum which was recently introduced in gravitational physics. Environmental effects, such as the addition of a thin accretion disk or a matter shell, lead to a secondary bump that appears in the effective potential of black hole perturbations. Regardless of the environment's small contribution to the effective potential, its presence can completely destabilize the fundamental quasinormal mode and may potentially affect black hole spectroscopy. Here, we perform a comprehensive analysis of such phenomenon for Schwarzschild, Reissner-Nordstr\"om, Schwarzschild-de Sitter, and Reissner-Nordstr\"om-de Sitter black holes by considering the potential for a test scalar field with the addition of a tiny bump sufficiently away from the photon sphere. We find a qualitatively similar destabilization pattern for photon sphere, complex, scalar quasinormal modes in all cases, and a surprising spectral stability for dominant scalar, purely imaginary, de Sitter and near-extremal modes that belong to different families of the spectrum. For Reissner-Nordstr\"om-de Sitter black holes, we re-evaluate the validity of the strong cosmic censorship and find that the addition of a realistic bump in the effective potential cannot prevent its violation due to a combination of the spectral stability of dominant de Sitter and near-extremal modes for small cosmological constants and an ineffective migration of the photon sphere modes that dominate the late-time ringdown signal for sufficiently large cosmological constants.
\end{abstract}

\maketitle

\section{Introduction}

The coalescence of two compact objects, such as black holes~(BHs) and neutron stars~(NSs), leads to the emission of gravitational waves~(GWs) that carry away significant information regarding the binary constituents. This provides numerous tests regarding the validity of General Relativity (GR)~\cite{LIGOScientific:2021sio}, the accelerated expansion of the Universe~\cite{LIGOScientific:2021aug}, the existence of new fundamental fields~\cite{Maselli:2021men}, and the properties of environments that surround binaries~\cite{Barausse:2014tra}, among many others~\cite{Barack:2018yly}. The triumphant detection of GWs from BH, NS and BH-NS mergers~\cite{LIGOScientific:2021djp} has turned the most cryptic aspects of the Universe from purely theoretical to observational. 

A fundamental aspect of the coalescence's remnant is its dynamical relaxation following the violent aftermath of the merger. The decay of perturbations in the exterior of BHs can be adequately described by quasinormal modes~(QNMs)~\cite{Kokkotas:1999bd,Berti:2009kk,Konoplya:2011qq}. BH spectroscopy is an important cornerstone of tests of gravity~\cite{Dreyer:2003bv,Berti:2005ys,Gossan:2011ha,Giesler:2019uxc,Isi:2019aib,Bhagwat:2019dtm,JimenezForteza:2020cve,Bhagwat:2021kwv,Forteza:2022tgq,Bhagwat:2023jwv,Baibhav:2023clw}, though the required level of modelling of these tests is still under scrutiny due to potential issues with the overtones and starting time~\cite{Cotesta:2022pci} and  with nonlinearities in the ringdown signal of BH mergers~\cite{Cheung:2022rbm,Mitman:2022qdl} (see~\cite{Baibhav:2023clw} for a recent detailed analysis).

On top of the aforementioned uncertainties regarding BH spectroscopy, further issues arise due to a delicate sensitivity of the QNM spectrum under tiny perturbations of the effective BH potential. BH QNMs undergo a \emph{spectral instability}; a phenomenon suggesting that when a tiny perturbation is added to the effective BH potential, the new QNM (or \emph{perturbed}) spectrum migrates in the complex plane disproportionately with respect to the scale of the added perturbation. Hints for such a spectral instability of Schwarzschild BHs were initially introduced in~\cite{Aguirregabiria:1996,Nollert:1996rf,Nollert:1998ys}, further explored in~\cite{Barausse:2014tra}\footnote{This effect is very much connected to the spectral instability against a change of the boundary conditions near the horizon, which is relevant in the context of echoes and exotic compact objects~\cite{Cardoso:2016rao,Cardoso:2019rvt,Maggio:2020jml,Maggio:2021ans}.}, and more recently discussed in~\cite{Jaramillo:2020tuu,Jaramillo:2021tmt,Gasperin:2021kfv,Jaramillo:2022kuv} through the notion of the \emph{pseudospectrum}~\cite{Trefethen:1993}. The rapidly expanding literature regarding spectral instabilities and the pseudospectrum of BHs~\cite{Jaramillo:2020tuu,Destounis:2021lum,Sarkar:2023rhp,Arean:2023ejh,alsheikh:tel-04116011} and horizonless compact objects~\cite{Boyanov:2022ark} indicates that QNM overtones are indeed spectrally unstable. Furthermore, it has been shown that including external matter to the BH, i.e. an accretion disk or simply a matter shell, can lead to not only the destabilization of overtones but also the fundamental frequency~\cite{Cheung:2021bol}. 
Although (at least) the fundamental mode extracted from the ringdown time-domain waveforms at early times seems to be spectrally stable against small perturbations of the effective BH potential~\cite{Nollert:1996rf,Nollert:1998ys,Barausse:2014tra,Berti:2022xfj}, overall the spectral instability complicates the applications of BH spectroscopy and tests of gravity~\cite{Franchini:2023eda,Destounis:2023ruj}, and calls for a more concrete consideration of astrophysical environments surrounding BHs~\cite{Barausse:2014tra,Cardoso:2021wlq,Cardoso:2022whc,Chen:2023akf}.

In Ref.~\cite{Cheung:2021bol} it was shown that even the fundamental quadrupole axial gravitational QNM can be destabilized by the addition of a minuscule bump on the exterior of the photon sphere~(PS) of a Schwarzschild BH. In particular, the migration of perturbed QNMs was shown to be amplified with the increase of the distance between the PS peak and the bump. Even though Schwarzschild BHs possess a quite straightforward QNM spectrum, that is associated with the angular frequency and decay timescale of null geodesics at the PS~\cite{Ferrari:1984zz,Cardoso:2008bp}, Reissner-Nordstr\"om~(RN) and asymptotically de Sitter~(dS) BHs have more intricate spectra that include more than one family of modes. In particular, RN BHs exhibit, besides PS QNMs, a family of near-extremal~(NE) modes which becomes dominant when the inner (Cauchy) and event horizons of such spacetimes approach each other~\cite{Richartz:2014jla,Hod:2017gvn}. On the other hand, Schwarzschild-de Sitter~(SdS) and Reissner-Nordstr\"om-de Sitter~(RNdS) BHs possess another set of modes, namely the de Sitter family, that are intrinsically connected to the positive cosmological constant and the cosmological horizon~\cite{Jansen:2017oag}. Therefore, RNdS BHs simultaneously admit three distinct families of modes: the dS, NE, and PS QNMs~\cite{Cardoso:2017soq}. 

One of the main scopes of this work is to perform a detailed analysis of QNM migration for all the aforementioned families of modes when environmental effects are taken into account, in a form of a matter-shell bump placed outside the PS. We consider a test scalar field on Schwarzschild, RN, SdS, and RNdS BHs, studying the migration of the fundamental QNMs by varying the bump parameters and exploring the rich parameter space of the background solutions.

Due to possible spectral instabilities of the aforementioned families, we also re-evaluate a fundamental aspect of GR, namely the strong cosmic censorship~(SCC) conjecture and its recently found violation in RNdS BHs~\cite{Cardoso:2017soq}. The SCC assesses that a Cauchy horizon, i.e. the boundary of maximal globally-hyperbolic development of initial data beyond which determinism is broken down~\cite{Hawking:1973uf}, becomes a mass inflation singularity under linear field perturbations~\cite{Ori:1991zz} due to a blueshift effect~\cite{Simpson:1973ua}. RN BHs~\cite{Dafermos:2012np}, indeed, do not violate the SCC~\cite{Dafermos:2003wr} due to an uneven clash between the inverse polynomial decay of linear perturbations in the exterior~\cite{Dafermos:2003yw,Dafermos:2010hb,Hintz:2020roc} and their exponential blueshift amplification at the Cauchy horizon~\cite{Dafermos:2003vim}. In dS BHs, though, linear perturbations in the exterior decay exponentially~\cite{Brady:1999wd,Molina:2003dc,Hintz:2016gwb,Hintz:2016jak,Konoplya:2022xid}, thus a violation is possible if the solutions of the linear perturbation equations are sufficiently regular at the Cauchy horizon~\cite{Hintz:2015jkj}. Recent studies regarding the  SCC~\cite{Cardoso:2017soq,Cardoso:2018nvb,Destounis:2019omd,Liu:2019lon,Mo:2018nnu,Dias:2018ufh,Davey:2022vyx,Casals:2020uxa,Destounis:2018qnb,Ge:2018vjq,Liu:2019rbq,Casals:2020uxa,Dias:2018etb,Dias:2018ynt,Destounis:2020yav} have resurfaced an interest in its potential violation as well as on how it can be quenched by adding matter fields akin to BH environments~\cite{Kehle:2021jsp,Nan:2023vkq}. Since the destabilization pattern of fundamental QNMs of Schwarzschild BHs~\cite{Cheung:2021bol}, predicted by the pseudospectrum~\cite{Jaramillo:2020tuu}, leads to larger decay timescales when adding a tiny bump outside the PS, due to stable trapping of waves between the PS peak and the bump, it may lead the blueshift effect at the Cauchy horizon of RNdS BHs to create a mass inflation singularity faster than the rate that the exterior damps linear perturbations, thus restoring the SCC. Hence, another of our main goals is to explore if the violation of the SCC in RNdS BHs surrounded by a matter shell still persists.

First, we explore the spectral instability of PS QNMs. We find that for all spacetimes considered, the PS family has always the tendency to migrate to smaller oscillatory frequencies and imaginary parts, as the bump is moved further away from the PS\footnote{Similar behavior has been found for dS BHs with position-independent, deterministic, perturbations introduced on the effective potential \cite{Sarkar:2023rhp}.}, making them more long-lived, after a transitory intermediate stage of outspiraling behavior, similar to that found in~\cite{Cheung:2021bol} for axial gravitational QNMs. The most significant effects are obtained for asymptotically-flat BHs. In this case, even though in the absence of a bump the large angular number $\ell\rightarrow\infty$ QNMs are the dominant ones (smallest imaginary part in absolute value), when the bump is sufficiently far from the BH potential the $\ell=0$ perturbed QNM overtakes the $\ell\rightarrow\infty$ mode. This is a rather interesting result, presented here for the first time, and is analogous to the anomalous decay rate of QNMs found in various BHs and theories of gravity~\cite{Lagos:2020oek,Aragon:2020tvq,Destounis:2020pjk,Aragon:2020xtm,Fontana:2020syy,Gonzalez:2022upu,Gonzalez:2022ote}. 

For asymptotically dS BHs, the migration of PS QNMs is suppressed significantly, due to the cosmological boundary imposed by the asymptotics of spacetime, but is still present when compared to the scale of the bump. Therefore, smaller cosmological constants, that have larger cosmological horizon radii, allow for further migration, since the bump can be placed further away from the PS without interfering with the cosmological boundary and the outgoing waves there, and vice versa. On the other hand, the dS modes of asymptotically-dS BHs are spectrally stable but their stability depends delicately to the location of the bump. When the bump is close to the PS then the migration is arbitrarily smaller with respect to the scale of the bump, since it does not interfere with the cosmological scales of the observable Universe. To the contrary, when the bump approaches the cosmological horizon then the dS modes of both SdS and RNdS spacetimes begin to show hints of spectral instability, since the bump begins to affect the asymptotic behavior of spacetime and its distance to the PS becomes comparable to the scale of the observable Universe. Nevertheless, even in these cases, the migration stays comparable to the overall scale of the perturbing bump, thus dS modes can still be considered spectrally stable when we refrain from changing the asymptotic structure of spacetime.

We further investigate the spectral stability of NE modes, when they are dominating the late-time ringdown signal, arbitrarily close to extremality in RN and RNdS BHs. We surprisingly find that this family, at least when dominant, is spectrally stable regardless to the bump's location. This phenomenon takes place no matter how close the bump is to the cosmological horizon, which designates a different underlying mechanism for their spectral stability. 

Finally, we re-evaluate the violation of the SCC for near-extremally charged RNdS BHs in regions of the parameter space where such violation occurs in the absence of a bump~\cite{Cardoso:2017soq}. Our  aforementioned analysis regarding the potential spectral (in)stability of all families of modes of RNdS BHs shows that the addition of a bump to the effective potential cannot prevent the violation of the SCC, due to an intricate combination of spectral stability of dominant dS modes for small cosmological constants and NE modes when they dominate close to extremal geometries, and an insufficient migration of dominant PS QNMs for large cosmological constants. Therefore, the results presented in~\cite{Cardoso:2017soq}, and recently discussed in~\cite{Yang:2022wlm}, are solid, at least for the kind of bump perturbation we have considered in this work.

In what follows we adopt geometrized units such that the gravitational constant $G$ and the speed of light $c$ equal to unity.

\section{Background spherically-symmetric BHs}\label{BHs}

In what follows, we consider static and spherically symmetric BHs that are either asymptotically flat or dS. In particular, the line elements we will explore have the form
\begin{equation}
    ds^2=-f_i(r)dt^2+f_i^{-1}(r)dr^2+r^2(d\theta^2+\sin^2\theta d\phi^2), \label{eq:bkg}
\end{equation}
where $f_i(r)$ are the respective lapse functions of Schwarzschild $f_\text{S}(r)$, RN $f_\text{RN}(r)$, SdS $f_\text{SdS}(r)$ and RNdS $f_\text{RNdS}(r)$ BHs, respectively, defined as
\begin{align}
    f_\text{S}(r)&=1-\frac{2M}{r},\label{S}\\
    f_\text{RN}(r)&=1-\frac{2M}{r}+\frac{Q^2}{r^2},\label{RN}\\
    f_\text{SdS}(r)&=1-\frac{2M}{r}-\frac{\Lambda}{3}r^2,\label{SdS}\\
    f_\text{RNdS}(r)&=1-\frac{2M}{r}+\frac{Q^2}{r^2}-\frac{\Lambda}{3}r^2,\label{RNdS}
\end{align}
with $M$, $Q$ and $\Lambda>0$ the mass, electric charge and cosmological constant of the aforementioned spacetimes~\cite{Hawking:1973uf}. All spacetimes involved possess a BH event horizon at $r=r_+$. In addition, the RN(dS) solution has an inner (Cauchy) horizon at $r=r_-<r_+$. Finally, asymptotically flat solutions such as the Schwarzschild and RN possess a null hypersurface at infinity, while asymptotically dS solutions, such as the SdS and RNdS, have a null hypersurface at $r=r_c$, namely the cosmological horizon, and a spacelike hypersurface at infinity. In these cases, the inequality $r_c>r_+$ must hold. For more details on the causal structure of the BHs discussed above we refer the reader to~\cite{Hawking:1973uf,Griffiths:2009dfa}.

\section{QNMs and their spectral stability}

To linear order in the scalar field perturbations, Einstein's equations decouple from the Klein-Gordon equation governing the evolution of the scalar field. This allows us to choose any GR solution, and we shall use the spherically symmetric spacetimes presented in the previous section as a fixed background where the linear scalar field propagates on. By choosing appropriate boundary conditions, one can study the modal stability of various spacetimes under such linear perturbations.

\subsection{Scalar QNMs}

The equation governing the evolution of linear massless neutral scalar fields on a fixed curved background reads
\begin{equation}\label{KG equation}
    \Box\Psi=\frac{1}{\sqrt{-g}}\partial_\mu\left(g^{\mu\nu}\sqrt{-g}\partial_\nu \Psi\right)=0,
\end{equation}
where $\Psi$ is the scalar field, $g_{\mu\nu}$ the metric tensor of spacetime, and $g=\det(g_{\mu\nu})$. Assuming the background~\eqref{eq:bkg}, decomposing the perturbation in spherical harmonics $Y_{\ell m}(\theta,\phi)$, and performing a transformation onto the Fourier domain, $\Psi=\sum_{lm}\frac{\psi(r)}{r}Y_{\ell m} e^{-i\omega t}$, Eq.~\eqref{KG equation} takes the form
\begin{equation}\label{master_equation}
    \frac{d^2\psi}{dr^2_*}+\left(\omega^2-V_i\right)\psi=0,
\end{equation}
where $dr/dr_*=f_i(r)$ is the tortoise coordinate, $\omega$ the eigenfrequency of the scalar field, and 
\begin{equation}\label{potentials}
    V_i=f_i(r)\left(\frac{\ell(\ell+1)}{r^2}+\frac{f_i^\prime(r)}{r}\right)
\end{equation}
is the effective potential. Here, prime denotes differentiation with respect to $r$ and the subscript $i$ denotes the different static and spherically symmetric backgrounds mentioned in the previous section, i.e. $i\in\{\text{S},\,\text{RN},\,\text{SdS},\,\text{RNdS}\}$ for Schwarzschild, RN, SdS and RNdS backgrounds, respectively. Assuming boundary conditions for the scalar field such that it describes purely ingoing waves at the event horizon and purely outgoing waves at infinity (or at the cosmological horizon for BHs with a positive cosmological constant), we obtain an infinite set of QNMs $\omega_{n \ell}$, where $n$ is the overtone number of the QNM such that $n=0$ denotes the fundamental mode that dominates the late-time ringdown signal, and higher $n$ denotes overtones. The QNMs of the potentials~\eqref{potentials} will be henceforth called the \emph{unperturbed QNMs} and will be denoted as $\omega_{n\ell}^{(0)}$.

As thoroughly explained in~\cite{Cardoso:2017soq}, the spectra of different BHs may exhibit a variety of families of modes, depending on the causal structure of spacetime. The omnipresent complex family of QNMs that all compact objects possess (besides some exotic exceptions~\cite{Destounis:2018utr,Konoplya:2017wot,Konoplya:2022gjp}) are an outcome of photon sphere (PS) excitations. The real and imaginary parts of this family are associated with the angular frequency and decay timescale of photons at the PS~\cite{Ferrari:1984zz,Cardoso:2008bp}. In near-extremally-charged RN(dS) BHs, a new family of NE modes appears, which are purely imaginary. These modes are proportional to the surface gravity of the Cauchy horizon\footnote{Or event horizon, since in the limit $Q\rightarrow Q_\text{max}$, with $Q_\text{max}$ the extremal charge, the Cauchy and event horizon coincide.}. Finally, asymptotically dS spacetimes display another family of purely imaginary modes associated with the cosmological horizon's surface gravity of pure dS space. In the limit where the cosmological constant vanishes, this family disappears.

\subsection{Perturbing the QNM spectrum}

To investigate the migration and possible destabilization of scalar QNMs of the BHs described in Sec.~\ref{BHs}, we assume a minuscule bump on the BH potentials~\eqref{potentials} that has the form of the P\"oschl-Teller function, also used in~\cite{Cheung:2021bol}, i.e.
\begin{equation}\label{perturbed_potential}
    V_\epsilon\equiv V_i+V_\text{bump},
\end{equation}
where 
\begin{equation}
    V_\text{bump}=\epsilon \sech^2(r-a)\,, \label{bump}
\end{equation}
and $a$ defines the bump position (not to be confused with the spin parameter of the Kerr BH). The strength of the perturbation is parametrized by the prefactor $\epsilon$ and the distance $a$. If we use $V_\epsilon$ instead of $V$ in the scalar wave equation~\eqref{master_equation}, and impose the same QNM boundary conditions, then we obtain the so-called \emph{perturbed QNMs} denoted as $\omega_{n\ell}^{(\epsilon)}$. In what follows, we will fix $\epsilon=10^{-6}$ and vary the dimensionless bump position $a/M$. The results for different $\epsilon$ can be qualitatively extrapolated from the phase diagrams in~\cite{Cheung:2021bol}.

Even though the addition of the bump ``by hand'' may seem as a rough estimate of an environment, it can arise from a mathematical procedure. It has been shown that the addition of a matter shell around a Schwarzschild~\cite{Cheung:2021bol} or RN~\cite{Feng:2022evy} BH can lead to an effective potential with the qualitative features of Eq.~\eqref{perturbed_potential}. Hence, astrophysical environments around BHs can lead to exact or numerical spacetime solutions that, when linear perturbations are considered, acquire an effective potential with a dominant PS peak and a highly subdominant bump (or negative well, see Appendix~\ref{AppA} for a detailed analysis) at some distance $a$ away from the PS, depending on the environment's configuration.

We are interested in the instability not only of fundamental scalar QNMs, but also of the modes from all families described above. We vary $a/M$ for all BHs presented in order to gain a complete understanding of mode migration in the complex plane, and whether this leads to a disproportional migration, designating a \emph{spectral instability}. On the other hand, if some mode does not migrate disproportionately (or at all) as the distance of the bump increases then this family shows signs of \emph{spectral stability} with respect to the bumpy perturbation imposed to the effective potential. To quantify the above, we define the difference between unperturbed and perturbed modes as $\Delta\omega^{(\epsilon)}\equiv |\omega^{(0)}-\omega^{(\epsilon)}|$ and normalize it as $|\Delta\omega^{(\epsilon)}/\omega^{(0)}|$, so that when the relative difference $|\Delta\omega^{(\epsilon)}/\omega^{(0)}|\gg\epsilon$ for a range of $a/M$ then a spectral instability occurs, otherwise (if $|\Delta\omega^{(\epsilon)}/\omega^{(0)}|\sim\epsilon$ or smaller) the particular family is spectrally stable. 

\subsection{Numerical method}

To compute the perturbed QNMs of a BH effective potential with a bump we use the shooting method~\cite{Pani:2013pma}, which relies on the direct integration of the master equation~\eqref{master_equation}. To do so, appropriate boundary conditions are chosen for QNMs and the field equations are solved analytically as a series expansion at both boundaries up to very high order, to ensure numerical accuracy~\cite{Pani:2013pma}. The scalar field's expansion at the boundaries, given in the definition of QNMs, is integrated from one boundary to a certain matching point, and then from the other boundary to the same matching point. Assuming continuity of solutions and their derivatives at the matching point we search for QNMs by checking if the matching condition is satisfied. For most of the parameter space scanned, the initial QNM guess is taken to be that for the case where $\epsilon=0$ and we iterate the code with the newly found migrated QNM as the new guess while $a/M$ increases, till we reach at least a minimum of $6$ digits of precision (in fact, most of the time the numerical precision is much higher).

\section{Perturbed spectra of asymptotically-flat spacetimes}

In this section we focus on the potential migration of QNMs of asymptotically-flat BHs. For clarity, we remind the reader that we solve the eigenvalue problem involved in Eq.~\eqref{master_equation} with the perturbed potential from Eq.~\eqref{perturbed_potential}, for Schwarzschild and RN geometries.

\subsection{Schwarzschild BHs}

\begin{figure}[t]
\centering
\includegraphics[scale=0.35]{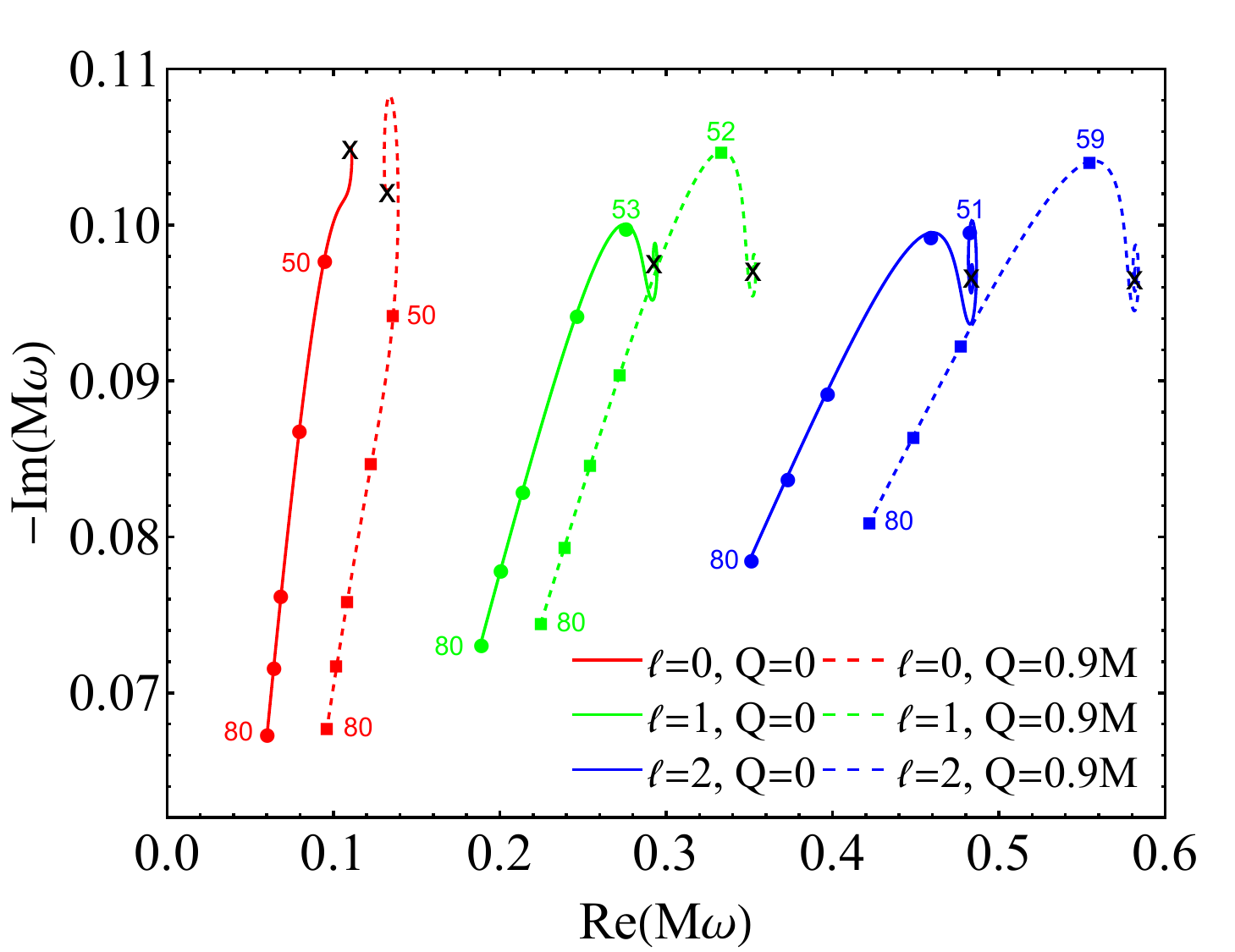}
\caption{PS scalar QNM migration parametrically shown in the complex plane as a function of $a/M$ (see values near data points) for Schwarzschild (solid curves) and RN BHs with $Q=0.9M$ (dashed curves) for $\ell=0,\,1,\,2$.}\label{PS_modes_flat}
\end{figure}

Schwarzschild BHs are the simplest spherically-symmetric background solutions considered in this work. Their unperturbed QNMs are characterized only by the effective potential peak near the PS, and its excitations. Thus, they only possess one family of QNMs, i.e. the PS QNMs. Adding a P\"oschl-Teller bump at some distance $a$ and continuously increasing $a$ away from the PS  leads to the solid migratory curves of Fig.~\ref{PS_modes_flat}. Since the bump scale is $\epsilon=10^{-6}$ and the migration of QNMs is $\mathcal{O}(10^{-2})$, the PS QNMs exhibit a spectral instability.

As discussed in the introduction, the dominant mode (i.e., the one with the smallest, in absolute value, imaginary part) of the unperturbed family of QNMs of Schwarzschild BHs is obtained in the geometric optics limit, $\ell\rightarrow\infty$, associated with unstable photon orbits at the PS. As $a/M$ increases, we observe different migrations of the QNMs for different values of $\ell$. The qualitative aspects of the migration due to the existence of the bump is to turn the perturbed QNMs less oscillatory (smaller real part) and more long-lived (larger decay timescales) after an outspiral trajectory (where the spectral instability already takes place) in the complex plane, moving them disproportionally away from their corresponding unperturbed QNMs. 

Interestingly, even though we only introduced the bump at a maximum distance from the event horizon $a/M=80$, the $\ell=0$ perturbed QNMs has overtaken the otherwise unperturbed fundamental $\ell=2$ QNM (and even its perturbed version). According to the pattern of migration in Fig.~\ref{PS_modes_flat}, which is quite similar to that of Ref.~\cite{Cheung:2021bol}, we can safely assume that the $\ell=0$ QNMs will always overtake the large $\ell$ mode (even if it is equally perturbed or not), thus leading to a similar effect such as the one found for massive scalar fields in Schwarzschild, i.e. the \emph{anomalous decay of QNMs}~\cite{Lagos:2020oek}, and for charged scalar fields in accelerating RN BHs when the acceleration parameters exceeds a threshold~\cite{Destounis:2020pjk,Destounis:2020yav}. In the aforementioned phenomena, there is a critical mass (or spacetime acceleration) beyond which the dominant mode of the PS family becomes the $\ell=0$ instead of the $\ell\rightarrow\infty$, in complete agreement with what occurs in our case for $a/M\gtrsim70$. Therefore, the addition of a bump in the BH's potential, is akin to adding a mass term in the scalar field, or in more general terms creating a trapping cavity that can support long-lived bound states. In a nutshell, we observe an overall spectral instability of the scalar PS QNMs in Schwarzschild BHs which is expected since even axial quadrupole QNMs demonstrate the same behavior~\cite{Cheung:2021bol}.

\subsection{RN BHs}

Figure~\ref{PS_modes_flat} shows also the migration of scalar PS QNMs of a RN BH with a rather large electric charge, $Q=0.9M$. As the bump distance is increased, the migration of RN PS QNMs is qualitatively similar to that of neutral PS ones: the PS modes become more long-lived and less oscillatory. Likewise, we observe the same effect of transition from dominant $\ell\rightarrow\infty$ to $\ell=0$ after $a/M\sim70$, in concordance with the anomalous decay rate of QNMs for massive scalar fields and its underlying mechanism.

\begin{figure}[t]
\centering
\includegraphics[scale=0.385]{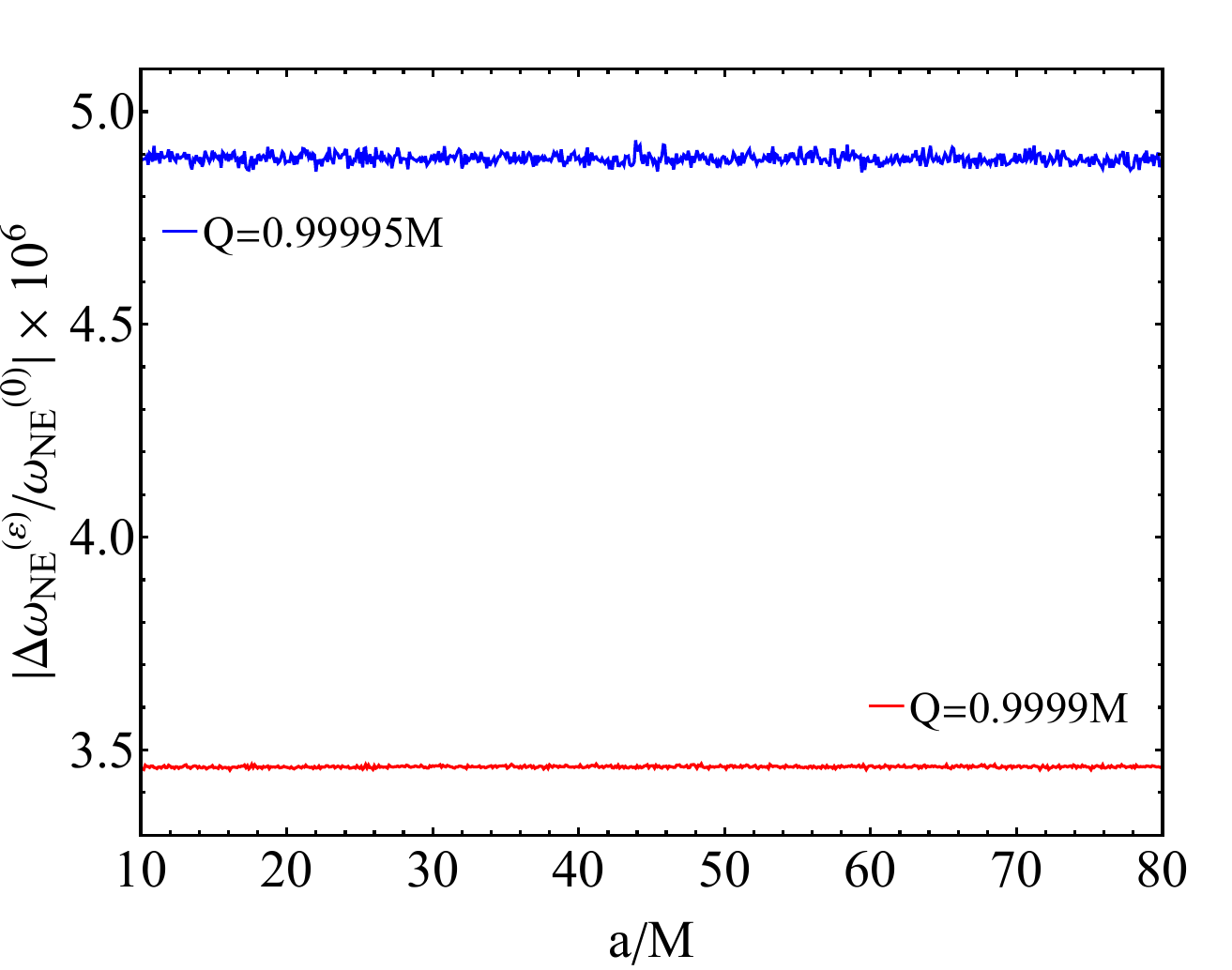}
\caption{Normalized migration of dominant normalized scalar NE $\ell=0$ modes of RN BHs as a function of $a/M$ for different values of the BH charge.}\label{NE_modes_RN}
\end{figure}

Besides the PS modes which are shown to be spectrally unstable, RN BHs possess another purely imaginary family of modes when $Q\rightarrow M$ for neutral linear scalar fields. These NE QNMs, also known as quasibound states~\cite{Rosa:2011my,Dolan:2015eua,Hod:2017gvn,Vieira:2021ozg,Vieira:2021xqw,Vieira:2023ylz}, are associated with the surface gravity at the Cauchy (or event) horizon arbitrarily close to extremality and dominate the ringdown signal in that regime~\cite{Cardoso:2017soq,Cardoso:2018nvb}. In Fig.~\ref{NE_modes_RN} we demonstrate that when this family is dominant, the corresponding fundamental mode is spectrally stable. Specifically, we considered two different RN BHs arbitrarily close to extremality, and found that the absolute normalized difference between the perturbed mode, as a function of $a/M$, and the unperturbed one ($a=0$) remains constant and of the same order as the bump scale. This result, presented to our knowledge for the first time here, can have important implications regarding the pseudospectra of NE RN BHs~\cite{Destounis:2021lum} and could be explained by the fact that NE eigenfunctions have support near the inner and event horizons (which approach each others in the dominant regime) and are insensitive to putative perturbations located near or outside the PS. Thus, these modes are not disproportionately affected by the type of  perturbations considered here.

\begin{figure}[t]
\centering
\includegraphics[scale=0.385]{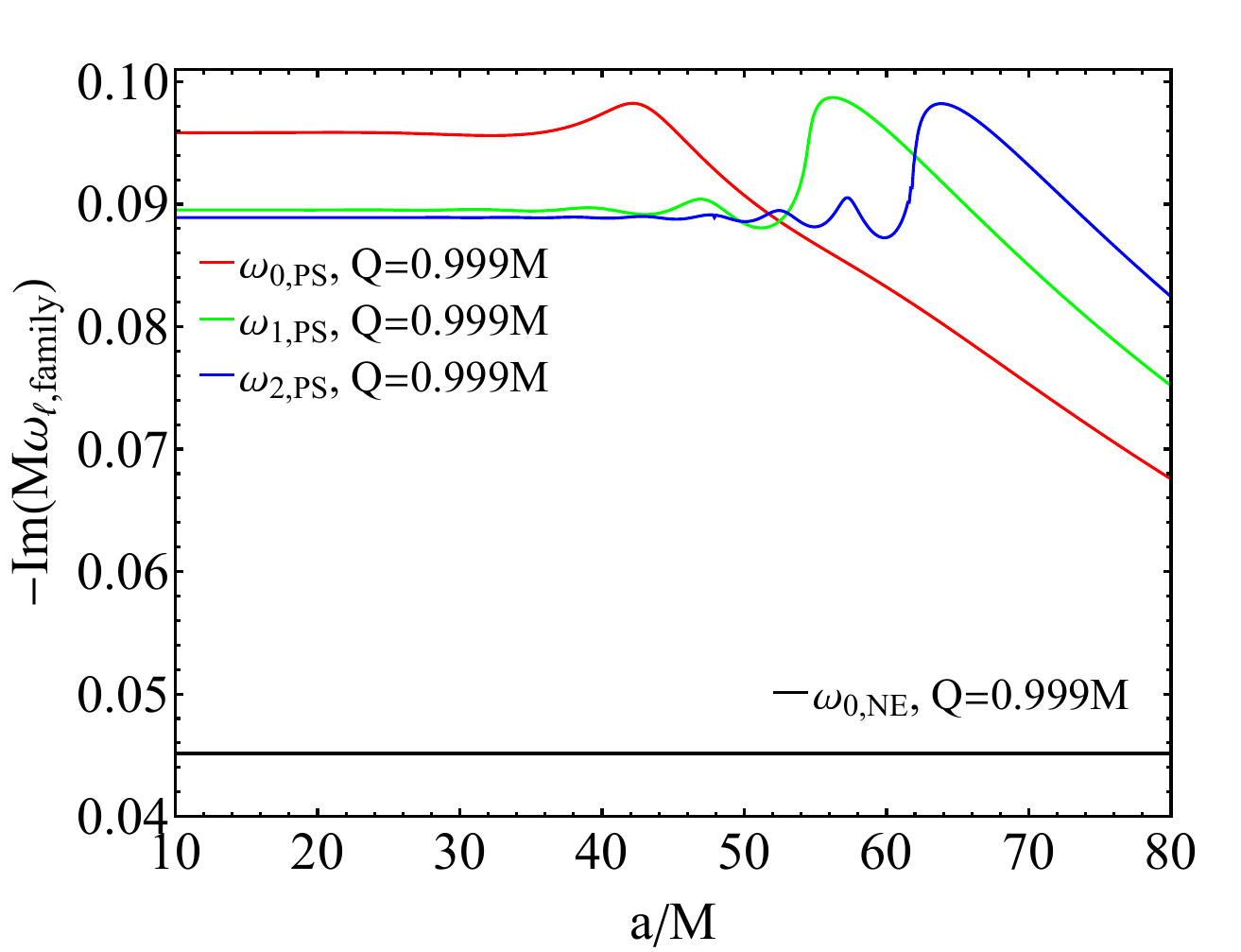}
\caption{Combined evolution of the $\ell=0,\,1,\,2$ PS scalar QNMs for a RN BH with $Q/M=0.999$, together with the dominant scalar NE mode with $\ell=0$.}\label{PS_NE_modes_vs_a}
\end{figure}

\begin{figure*}[t]
\includegraphics[scale=0.33]{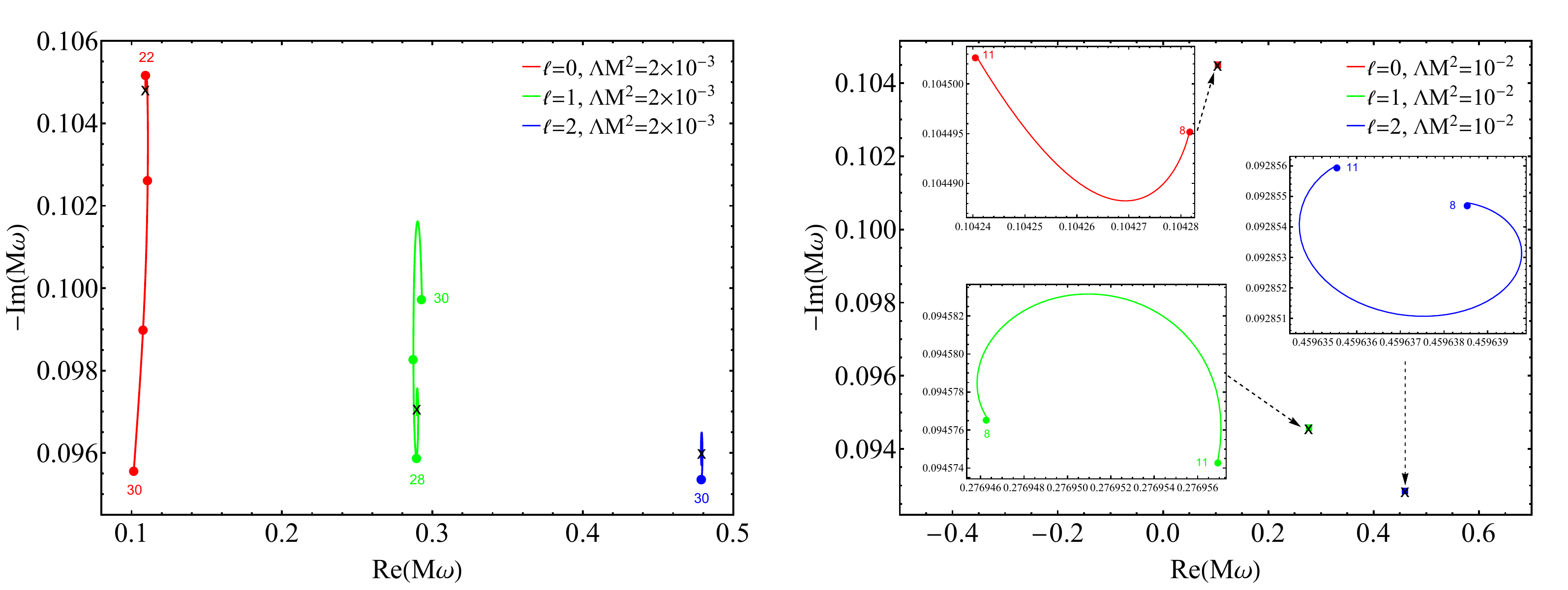}
\caption{Left: Scalar PS QNM migration as a function of $a/M$ for a SdS BH with a cosmological constant $\Lambda M^2=2\times 10^{-3}$. Right: Same as left but with $\Lambda M^2=10^{-2}$. In this case migration is much more limited and the three insets are zoom-ins of the evolution of $\ell=0,1,2$ modes.
}\label{SdS_PS_modes}
\end{figure*}

To summarize the spectral (in)stability of QNMs in RN geometries, in Fig.~\ref{PS_NE_modes_vs_a} we present both families of modes for a charge close to extremality, where NE modes are relevant. As $a/M$ increases, we observe that the imaginary parts of PS QNMs have a quasi-stable phase until they start oscillating and then migrate disproportionately (to order $\mathcal{O}(10^{-2})$) with respect to the scale of the perturbing bump (see Fig.~\ref{PS_modes_flat}). On the other hand, the dominant NE mode appears to be perfectly stable and displays a clean plateau, regardless of the choice of $a/M$. This is a strong indication of spectral stability with respect to the particular perturbation bump used in our analysis. Despite the fact that we cannot extend Fig.~\ref{PS_NE_modes_vs_a} for larger values of $a/M$ due to numerical limitations, an extrapolation of the migration of the PS modes to larger $a/M$ suggests that it would be the $\ell=0$ PS mode (or some of the PS QNM overtones~\cite{Cheung:2021bol}) that would eventually become dominant at sufficiently large $a/M$, superseding the stable NE mode.

\section{Perturbed spectra of asymptotically-dS spacetimes}

In this section we examine the migration of QNMs of BHs residing in a Universe with a positive cosmological constant. In what follows, we solve Eq.~\eqref{master_equation} with the potential in Eq.~\eqref{perturbed_potential} and appropriately outgoing boundary conditions at the cosmological horizon. The existence of the latter will become crucial to the enrichment of the spectrum of SdS and RNdS BHs under linear neutral scalar fields.

\subsection{SdS BHs}

As thoroughly discussed, SdS BHs possess, besides an event horizon, another null hypersurface, namely the cosmological horizon. Beyond the cosmological horizon incoming waves are infinitely redshifted due to the accelerated expansion of the Universe, therefore the observable part of the Universe is between the event and cosmological horizon radii. 

In this finite region, we find two families of modes in the spectrum, namely the PS modes~\cite{Cardoso:2008bp} and the dS modes~\cite{Lopez-Ortega:2006aal,Jansen:2017oag,Cardoso:2017soq}. As in asymptotically-flat BHs, the PS QNMs are prone to spectral instabilities when a P\"oschl-Teller bump is introduced to the effective potential. On the left panel of Fig.~\ref{SdS_PS_modes}, a qualitatively similar effect is found as the one in Fig.~\ref{PS_modes_flat}. Nevertheless, the migration is quantitatively different from that of Schwarzschild BHs due to the existence of the cosmological horizon. 

On the left panel of Fig.~\ref{SdS_PS_modes}, the $\ell=0,\,1,\,2$ PS QNMs migrate enough to be characterized as spectrally unstable, with the smallest migration being of order $\mathcal{O}(10^{-3})$ for $\ell=2$. This is visible and can occur due to the rather small choice of $\Lambda M^2=2\times 10^{-3}$ . The reasoning behind this choice of the cosmological constant\footnote{Unfortunately, our numerical scheme does not allow us to use smaller values of the cosmological constant due to numerical inaccuracies. Nevertheless, a smaller cosmological constant leads to a larger cosmological horizon, and we do expect an even stronger spectral migration of PS QNMs with respect to the bump's scale.} is that the smaller it is, the further the cosmological horizon radius is and therefore we can place the bump further from the PS. If on the other hand we use a rather large cosmological constant, such as the one on the right panel of Fig.~\ref{SdS_PS_modes} ($\Lambda M^2=10^{-2}$), the migration of modes is minimal but still at least an order of magnitude larger (for the $\ell=0$ QNM) than the scale of the bump scale. This occurs due to the fact that the observable Universe becomes much smaller (or the SdS BHs becomes large enough) and therefore there is not enough radial range to place the bump, thus the migration is minimized. 

To sum up, smaller cosmological constants will allow for a larger and disproportionate PS QNM migration in the complex plane, which triggers a spectral instability in SdS BHs. To the contrary, large cosmological constants minimize and eventually counterbalance the PS QNM migration due to bump positioning minimization. We note that this only occurs for the particular perturbation we considered in the effective potential that is position-dependent. Other perturbations such as ``sinusoidal'' or ``random''~\cite{Jaramillo:2020tuu,Jaramillo:2021tmt,Jaramillo:2022kuv}, that are not position-dependent, as well as ``cuts'' to the asymptotic region of the potential~\cite{Nollert:1996rf,Daghigh:2020jyk,Shen:2022xdp} that are position-dependent, or slightly perturbed different boundary conditions~\cite{Maggio:2020jml,Chakraborty:2022zlq} may destabilize the PS QNMs of SdS BHs at observational levels even for large cosmological constants.

\begin{figure}[t]
\centering
\includegraphics[scale=0.36]{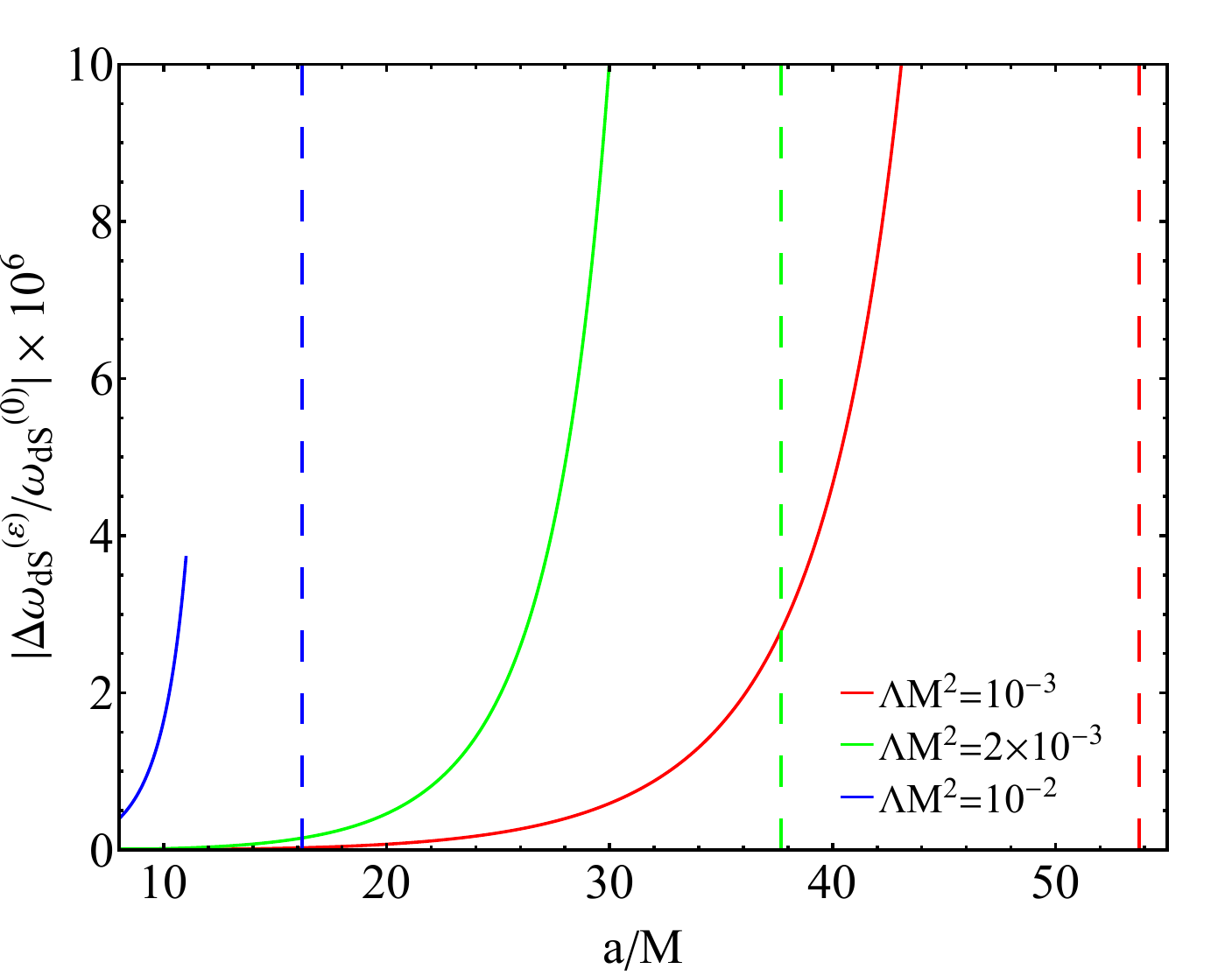}
\caption{Evolution of the dominant normalized scalar dS mode with $\ell=1$ of SdS BHs as a function of $a/M$ and for different values of the cosmological constant. The vertical dashed lines denote to the cosmological horizon radii corresponding to  the different choices of $\Lambda M^2$.}\label{SdS_dS_modes}
\end{figure}

The second family of modes that exist in dS BHs and their migration with respect to the bump location is presented in Fig.~\ref{SdS_dS_modes}. This figure is demonstrating complementary information regarding the insufficient migration of PS QNMs when the cosmological constant is sufficiently large. We observe that: (i) the dS modes are stable up to a point where they start diverging close to the cosmological horizon. This occurs because the closer the bump is located to the cosmological horizon, the more it interferes with the asymptotic behavior of spacetime and the stable trapping region created between the PS and the bump acquires scales similar to that of the observable Universe.
%
\begin{figure}[t]
\centering
\includegraphics[scale=0.35]{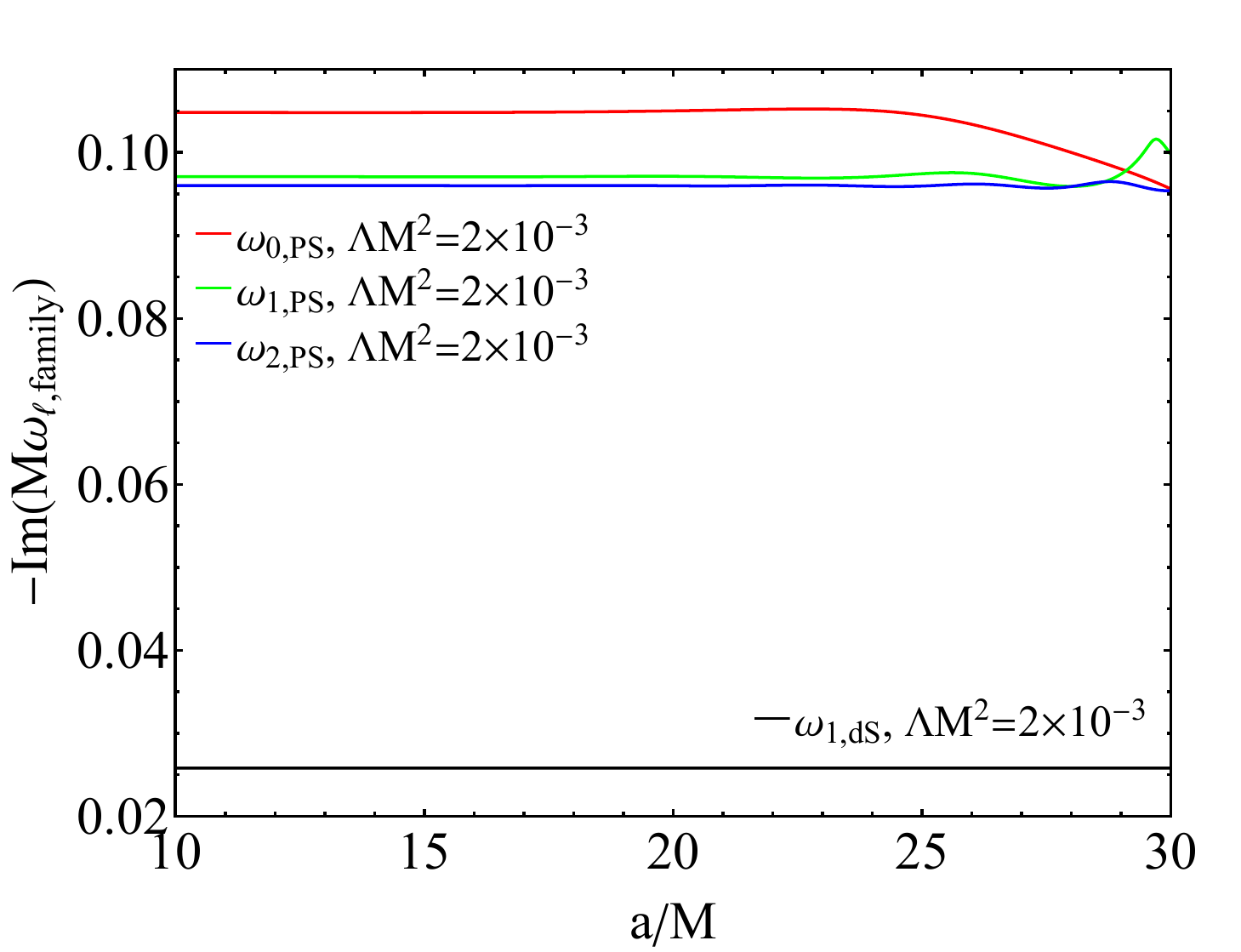}
\caption{Combined evolution of the $\ell=0,\,1,\,2$ photon sphere scalar QNMs of a SdS BH with $\Lambda=2\times 10^{-3}$, together with the dominant scalar de Sitter mode with $\ell=1$.}\label{SdS_PS_dS_modes_vs_a}
\end{figure}
%
(ii) The smaller the cosmological constant is, the larger the position of the bump can be where spectral stability occurs. Since the cosmological constant in our Universe is actually extremely small, we expect that the dS modes are spectrally stable for all practical purposes. To our knowledge, this is the first time that the spectral stability of yet another family of mode appears in the literature, beside the stability of NE mode in RN BHs discussed in previous sections.

For completeness, we also show the combined evolution of the PS QNMs together with the dominant dS mode for a fixed cosmological constant, $\Lambda=2\times 10^{-3}$ (see Fig.~\ref{SdS_PS_dS_modes_vs_a}). As in RN BHs, the PS modes eventually are destabilized and migrate for at least three order of magnitude more than the bump's scale, though in a different fashion compared to the asymptotically flat case,  due to the restricted range of the bump position. In contrast, the dS mode remains spectrally stable, at least for our choice of $\epsilon$. Nevertheless, even if we increased $\epsilon$, there will always be a smaller cosmological constant that stabilizes the dS modes, for choices of $a/M$ that do not interfere with the asymptotic behavior of the BH potential.

\subsection{RNdS BHs}

In the last section of our spectral stability analysis we will examine linear scalar field QNM migration in RNdS BHs, that possess all three aforementioned families of modes in their spectra. For the rest of this analysis, and according to the previous sections, we will make an educated choice for the cosmological constant, i.e. $\Lambda=2\times 10^{-3}$ and vary the BH charge $Q/Q_\text{max}$. This choice ensures that there is enough radial range for the the bump location and, in addition, it selects a range in which our numerical scheme remains highly accurate.

\begin{figure}[t]
\centering
\includegraphics[scale=0.34]{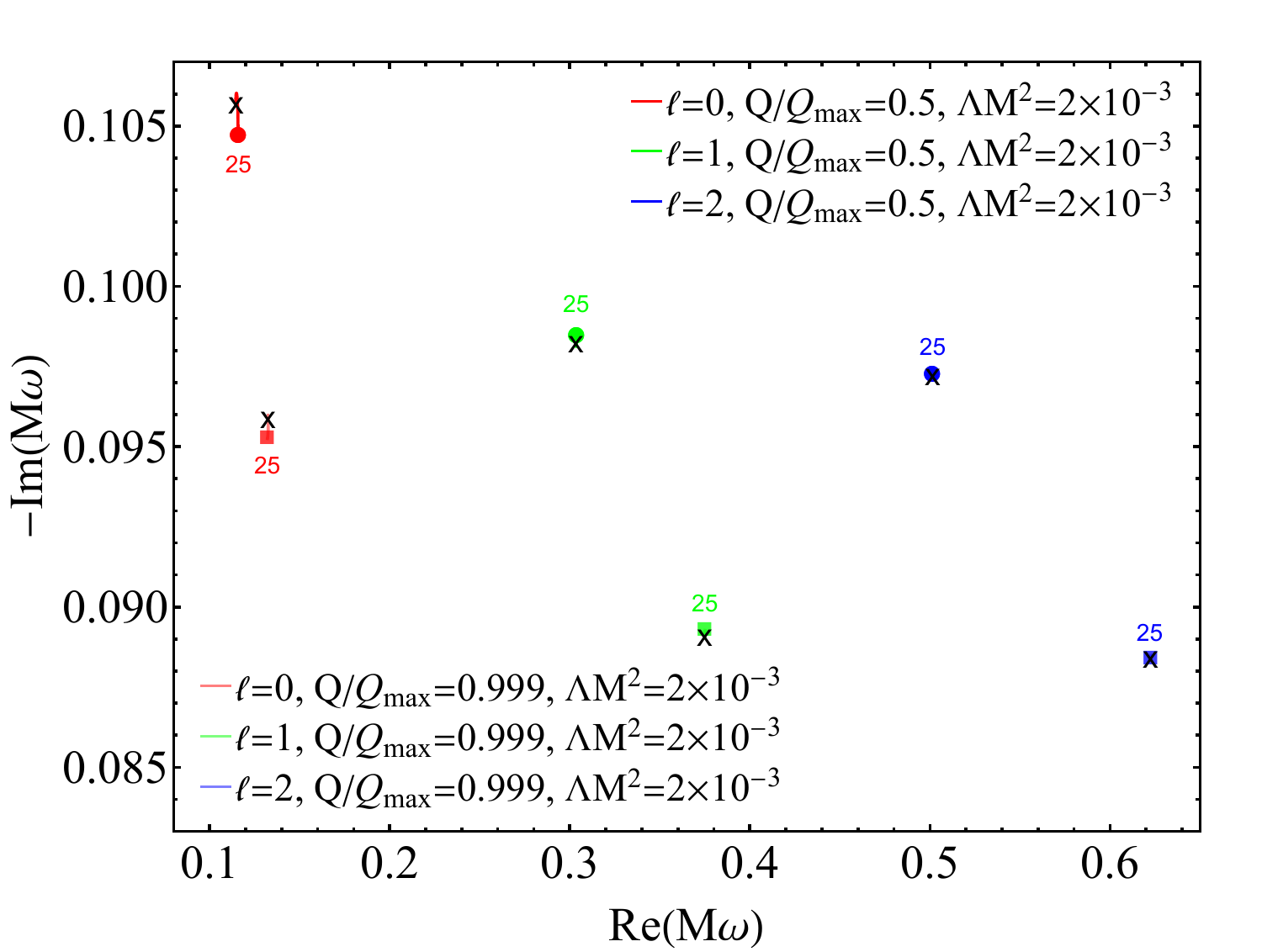}
\caption{Scalar PS mode migration in the complex plane with respect to the position of the bump. The crosses in all cases designate the unperturbed PS QNMs of a RNdS BH with $\Lambda M^2=2\times 10^{-3}$ and varying $Q/Q_\text{max}$. The colored (lightly-colored) curve designate the migration trajectory for $Q/Q_\text{max}=0.5$ ($Q/Q_\text{max}=0.999$.)}\label{RNdS_PS_modes}
\end{figure}

Figure~\ref{RNdS_PS_modes} shows the migration of PS QNMs of RNdS spacetime. Even though the cosmological horizon stays approximately the same with respect to the SdS case (since it depends strongly on the cosmological constant choice and only mildly on the charge), the inclusion of charge makes the migration even more minimal due to the shrinking of the available range for the bump location where our numerical integration is accurate and trustworthy. Even though Fig.~\ref{RNdS_PS_modes} looks pessimistic in what regards a potential spectral instability of PS QNMs, the migration for (at least) $\ell=0$ PS QNM is three order of magnitude larger than the scale of the bump. Therefore, the PS modes are still spectrally unstable, especially if we consider smaller cosmological constants. Crucially, the larger the charge, the more dominant (i.e., longer lived) the QNMs become, while due to their insufficient migration, the dominant QNM of the PS family, even with the addition of the bump, still remains the $\ell\rightarrow\infty$, in contrast to Schwarzschild and RN BHs.

\begin{figure}
\centering
\includegraphics[scale=0.34]{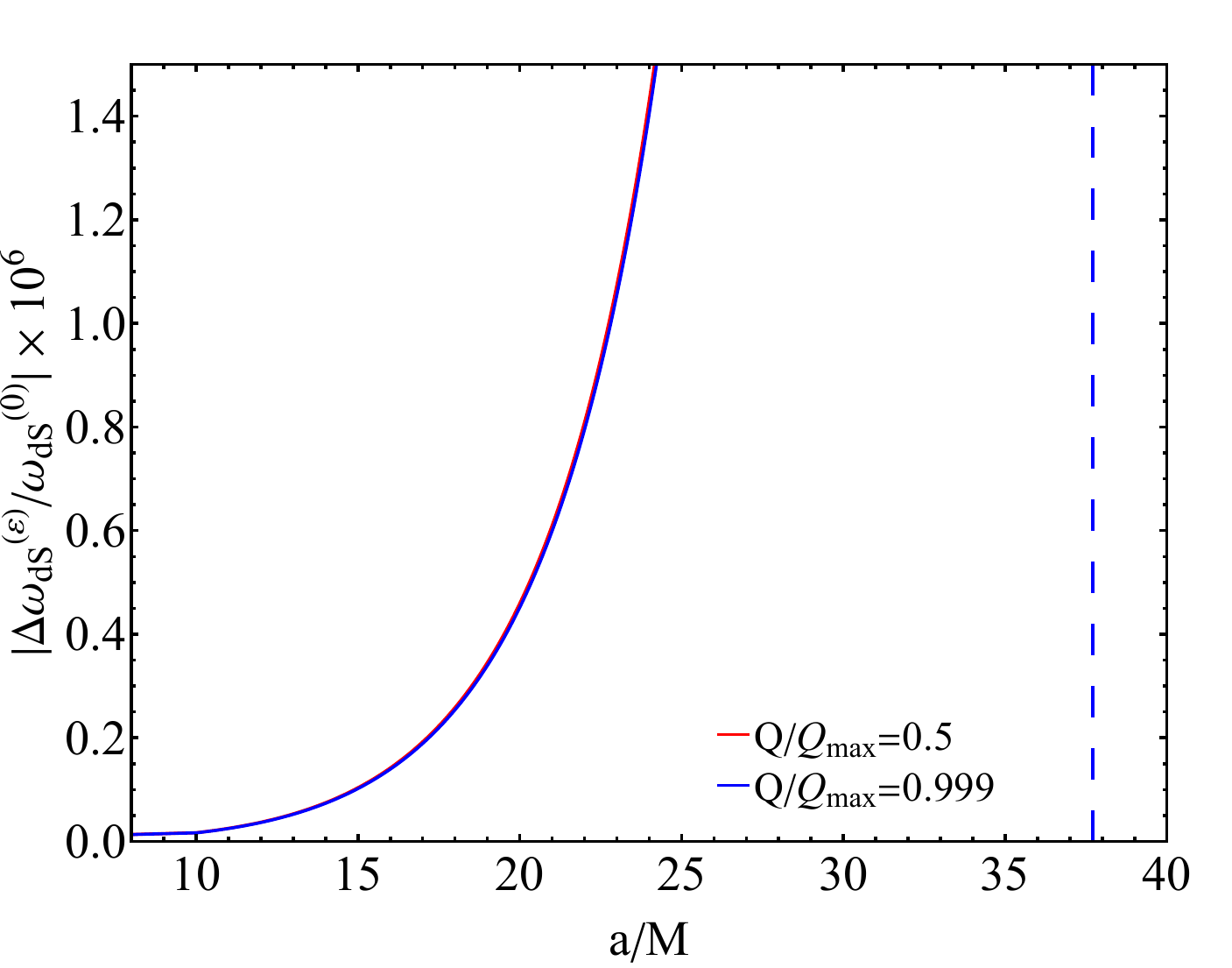}
\caption{Evolution of the normalized difference between the unperturbed and the perturbed dS modes of a RNdS BH with $\Lambda M^2=2\times 10^{-3}$ and for two different values of the charge. The vertical dashed lines (which almost overlap) designate the cosmological horizon radii for the two BH cases with different charges.}\label{RNdS_dS_modes}
\end{figure}

\begin{figure}
\centering
\includegraphics[scale=0.38]{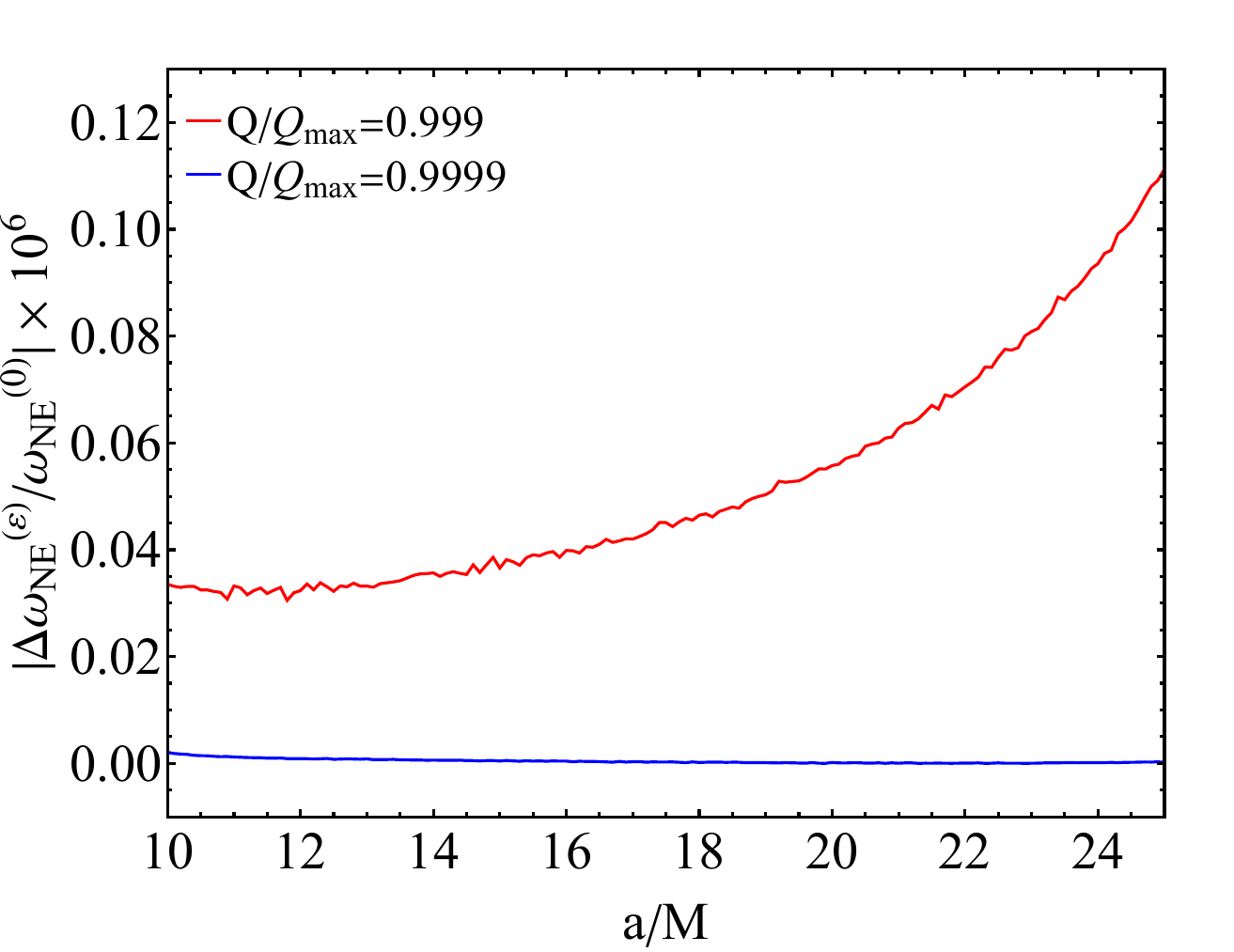}
\caption{Evolution of the normalized difference between the unperturbed and the perturbed NE modes of a RNdS BH with $\Lambda M^2=2\times 10^{-3}$ and for two different values of the charge.}\label{RNdS_NE_modes}
\end{figure}

Figure~\ref{RNdS_dS_modes} shows the migration of the dS modes with respect to $a/M$ for two different BH charges; a moderate value and another one close to extremality. Nevertheless, their trend is practically indistinguishable, in accord to~\cite{Cardoso:2017soq}; the dS modes are oblivious to the BHs' characteristics. They practically depend only with the cosmological horizon radius, which changes drastically only when the cosmological constant varies. 

\begin{figure}[h!]
\centering
\includegraphics[scale=0.38]{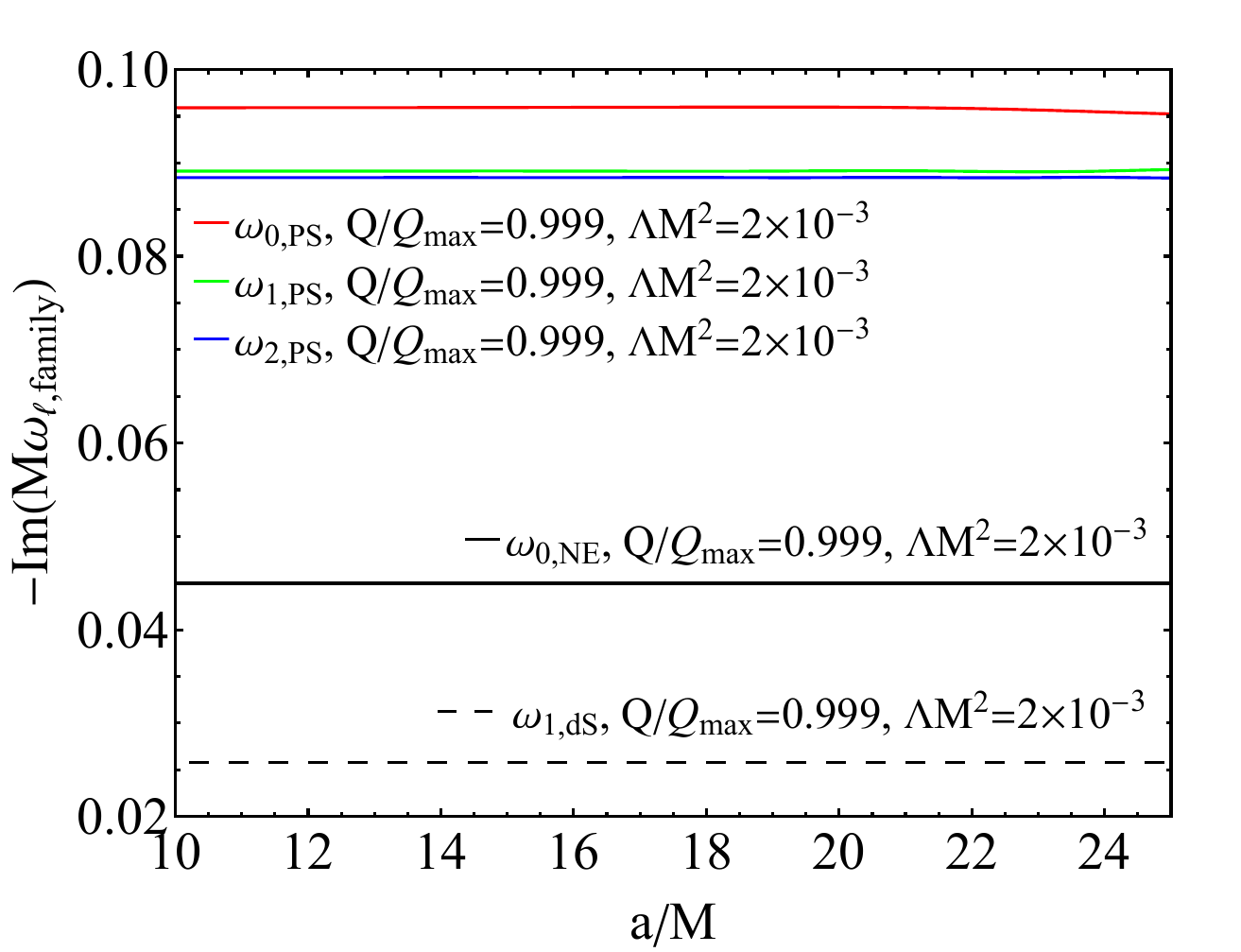}
\caption{Evolution of the imaginary part of all families of QNMs of a RNdS BH with $\Lambda M^2=2\times 10^{-3}$ and $Q/Q_\text{max}=0.999$. In this case, the $\ell=1$ dS family is the dominant one shown with a dashed horizontal line.}\label{RNdS_all_modes}
\end{figure}

The pattern in Fig.~\ref{RNdS_dS_modes} is qualitatively similar to that found for the dS mode migration in SdS BHs (see Fig.~\ref{SdS_dS_modes}) and designates spectral stability, as long as we do not place the bump sufficiently close to the cosmological horizon, that drastically changes the asymptotic form of the effective potential. Nevertheless, smaller values of the cosmological constant (closer to its actual measured value) allow for a more prolonged range of spectral stability with respect to $a/M$, hence we should be able to always counterbalance the region of spectral instabilities, close to the cosmological horizon, and keep the dS modes spectrally stable.

Due to the existence of charge, RNdS BHs also exhibit NE modes close to extremality. In Fig.~\ref{RNdS_NE_modes} we demonstrate that the NE modes are spectrally stable against a bump of order $\epsilon=10^{-6}$. Especially when the charge is almost extremal ($Q/Q_\text{max}=0.9999$) the normalized difference of the unperturbed with respect to the perturbed mode is practically zero no matter the bump's position since for this case the NE mode is the dominant one. Nevertheless, when the BH is close to extremality, and the corresponding modes are spectrally stable, for choices where the dS mode dominates the late stage of the ringdown (see, e.g., Fig.~\ref{RNdS_all_modes} for $Q=0.999 Q_{\rm max}$), the NE overtone mode has a tendency to grow (see e.g. the red curve in Fig.~\ref{RNdS_NE_modes}) and if the cosmological constant was even smaller there might be a chance that this particular overtone could be mildly destabilized. Thus, we may conjecture that when there are two competing imaginary modes, the obvious spectrally stable one is the dominant one while other imaginary overtones may get destabilized but only for very small cosmological constants. As far as we have checked, we have not been able to find such a case though, thus we can assume that NE modes of RNdS BHs are spectrally stable when the effective potential is perturbed by a P\"oschl-Teller bump. For completeness, in Fig.~\ref{RNdS_all_modes}, we present all modes with respect to the bump's position for a crucial case for the upcoming section. This figure clearly shows that there is no `mode mixing', i.e., the dominant dS mode does not cross at any point the other two families, i.e. the NE and PS overtones.

\section{Revisiting the SCC conjecture}

The previous sections suggest the following conclusions: (i) PS QNMs of asymptotically flat neutral and charged BHs can disproportionally migrate in the complex plane when a minuscule bump is added to the effective potential. For these spacetimes, the migration of $\ell=0$ PS QNMs is so sharp that it can overtake the $\ell\rightarrow\infty$ QNMs which are dominant for $\epsilon=0$. This is a strong indication of spectral instability. (ii) PS QNMs of asymptotically-dS, neutral and charged BHs migrate much less due to the existence of the cosmological horizon and the bound it imposes on the maximum position that the bump can have without interfering with the and the asymptotic form of the effective potential and eventually the boundary conditions. Nevertheless, their minimal migration still indicates spectral instabilities. (iii) Purely imaginary dS and NE modes are spectrally stable under the addition of a tiny bump in a large portion of the observable Universe. Both families are oblivious to perturbations of the effective potential, no matter the asymptotics of spacetime being flat or dS. Owing to the migration of PS modes and the spectral stability of the rest of the families, it is relevant to understand which is the mode that dominates the ringdown at late times in a certain parameter space and whether its decay timescale can be relevant in the context of the SCC. 

The modern formulation of the SCC conjectured by Christodoulou~\cite{Christodoulou:2008nj} states that \emph{the maximal globally-hyperbolic development of generic initial data are future inextendible beyond the Cauchy horizon as a spacetime with square-integrable Christoffel symbols}. This conjecture practically breaks down into two equally-important parts. The first part involves the fact that the Einstein field equations should not admit any solution beyond the Cauchy horizon\footnote{The boundary of maximal globally-hyperbolic development of generic initial data.}, not even a \emph{weak solution}, that is not exact with continuous second derivatives of the metric and matter fields involved, but rather a ``solution'' of the field equations that when integrated on both sides has enough regularity for extensions to exist, such as shock waves im hydrodynamics~\cite{Gourgoulhon:2006bn} and GR~\cite{Papapetrou:1975kj,Ishizuka:1980,Barrabes:2011lub}. This part, therefore, defines the regularity condition for extendibility beyond the Cauchy horizon. The second part, that is related to the probe of the SCC with the scalar field wave equation at the Cauchy horizon, translates the regularity condition of extensions beyond it to a mathematical inequality which decides if SCC is violated or not. It turns out (see Appendix~\ref{beta_proof} and Section 5.4 in Ref.~\cite{Destounis:2019zgi}) that this inequality combines the late-time exponential decay of scalar perturbations at the cosmological horizon, i.e. the fundamental/lowest-lying QNM of all families of RNdS spacetimes, and the strength of the exponential blueshift at the Cauchy horizon which is related to the surface gravity at the Cauchy horizon. As detailed in Appendix~\ref{beta_proof}, the inequality for vacuum and non-vacuum RNdS geometries, has the form 
\begin{equation}
    \beta\equiv\text{inf}(-\text{Im}(\omega_{n\ell}))/\kappa_->1/2\,, \label{beta}
\end{equation}
where $\kappa_-=-f^\prime_\text{RNdS}(r)/2|_{r=r_-}=f^\prime_\text{RNdS}(r_-)/2$ is the surface gravity of the Cauchy horizon of a RNdS BH.

Indeed, Ref.~\cite{Cardoso:2017soq} studied the SCC for neutral massless scalar fields in RNdS BHs to search for potential violations and found that the SCC is violated ($\beta>1/2)$ near extremality, in a significant volume of the available parameter space of the BH, thus allowing for potential observers to cross it smoothly enough, without experiencing infinite tidal deformations~\cite{Ori:1991zz}, and violate both the deterministic nature of the field equations and the SCC. Therefore, we can perform a similar analysis of the SCC and its potential violation when a matter shell bump is present in the effective potential of RNdS BHs, with the hope that the PS QNM spectral instability can migrate enough in order to minimize (or possibly rule out) the parameter space of the BH where such violation occurs.

\begin{figure}[t]
\centering
\includegraphics[scale=0.34]{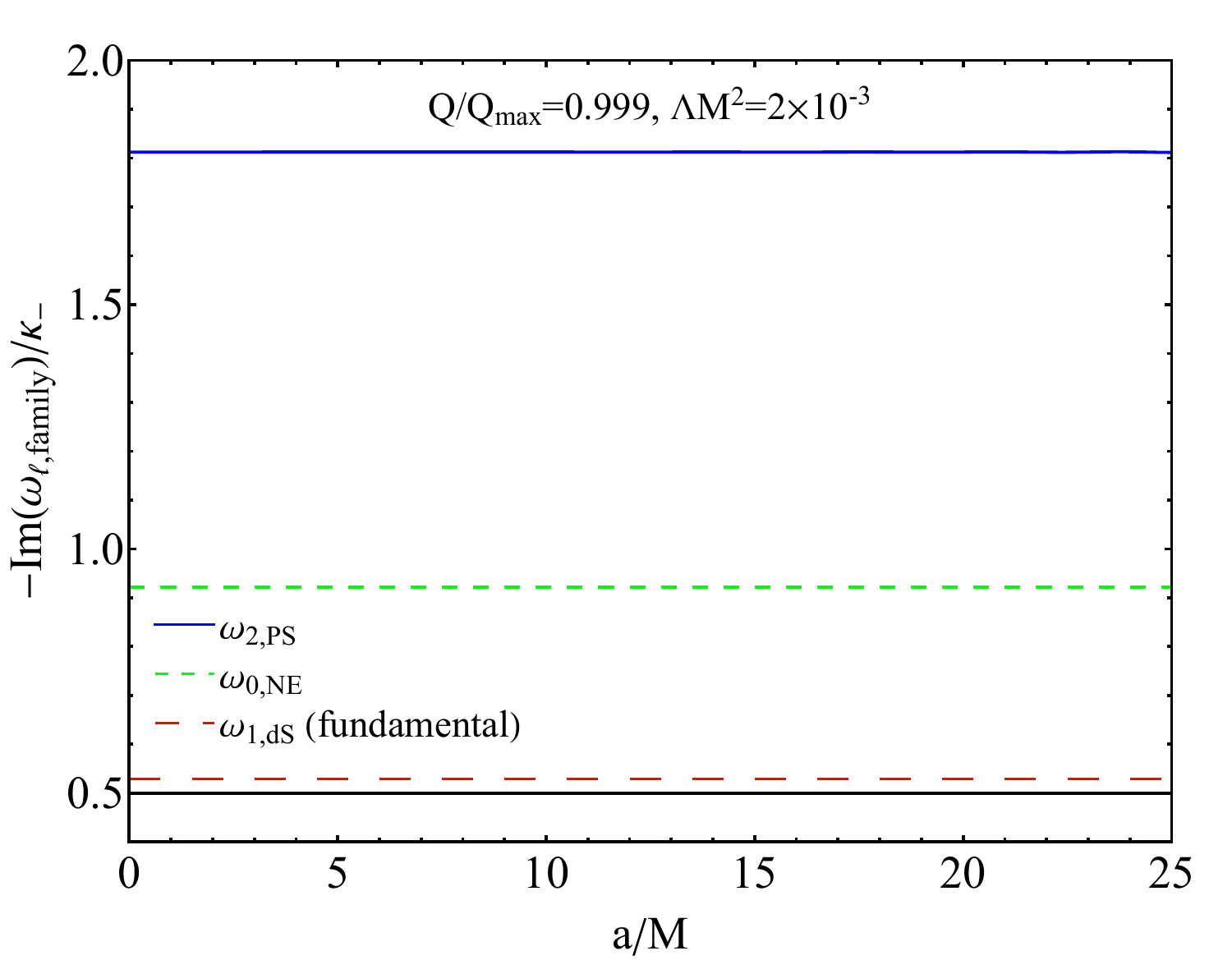}
\caption{Imaginary part of dominant modes normalized by the surface gravity of the Cauchy horizon a RNdS BH with $\Lambda M^2=2\times 10^{-3}$ and $Q/Q_\text{max}=0.999$. Differently-colored lines designate different dominant modes from all families present in the spectrum. Here, we show the (mild) migration of the (nearly) dominant $\ell=2$ PS QNMs (shown in blue), the dominant $\ell=1$ dS mode (shown in hardly dashed red) and the dominant $\ell=0$ NE mode (shown with mildly dashed green) with respect to $a/M$. The black horizontal line designates $\beta=1/2$, beyond which the SCC is violated.}\label{fig:beta}
\end{figure}

Our case study focuses on a RNdS BH with $\Lambda M^2=2\times 10^{-3}$, $Q/Q_\text{max}=0.999$ and a perturbed effective potential with a P\"oschl-Teller bump of amplitude $\epsilon=10^{-6}$ that leads to spectral instabilities in the PS QNMs. For the particular case study, and assuming that the bump is absent, SCC is violated since all dominant modes from all families satisfy $\beta>1/2$. More precisely, $\beta=\text{inf}(-\text{Im}(\omega_{n\ell}))/\kappa_-= 0.529$. Here, the dominant mode of all families for this particular spacetime is the $\ell=1$ dS fundamental mode, due to the choice of the cosmological constant. Therefore, the BH charge $Q/Q_\text{max}=0.999$ is close to the value above which violation occurs.

Figure~\ref{fig:beta}, demonstrates how the quantity $-\text{Im}(\omega_{\ell,\text{family}})/\kappa_-$ changes as the bump position $a/M$ increases, where we have set $n=0$ to only obtain the fundamental mode of each family. Due to our extensive analysis of spectral migration in the previous sections for RNdS BH spectra, we observe that all families of modes satisfy $-\text{Im}(\omega)/\kappa_->1/2$, no matter the choice of $a/M$, in the allowed region that does not affect the asymptotic form of the potential (see Figs. \ref{RNdS_dS_modes} and \ref{RNdS_all_modes}). This is due to an insufficient migration of the fundamental PS QNM with large $\ell$ (we roughly approximate the large $\ell$ limit of the perturbed QNMs with $\ell=2$) and the spectral stability of the dominant mode of the dS ($\ell=1$) and NE ($\ell=0$) families. For this particular BH, the dominant mode among all families is the dS one which lies close to $\beta=1/2$ but slightly exceeds it. Hence, we can conclude that $\beta$ corresponds to the dashed red line in Fig.~\ref{fig:beta}, i.e. the lowest-lying mode that will define the numerator in Eq.~\eqref{beta} and survive longer during the ringdown. Therefore, the SCC is still violated for small cosmological constants, regardless of the spectral instability of PS QNMs. For large cosmological constants the results should be the same due to an even more insufficient migration of the PS QNMs near extremality, where the SCC is violated. 

Nevertheless, we cannot exclude mode crossing, i.e. the migratory fundamental QNM of the PS family taking over the dominant mode, for much smaller $\Lambda$ than the one considered. If indeed, the PS QNM can migrate enough and overtake all other families, it may lead to $\beta<1/2$ which can effectively heal SCC. Nonetheless, we expect that this could only occur in a rather small portion of the available parameter space where the cosmological constant is very small and $a/M$ is unrealistically large. Even in this regime, the dS mode could still be the dominant one, regardless of the PS QNM migration, since it is proportional to the surface gravity of the cosmological horizon of pure dS space that tends to zero as $\Lambda$ decreases\footnote{For example, a cosmological constant $\Lambda=10^{-20}$ leads to a fundamental $\ell=1$ mode of the dS family of order $\omega_\text{dS}\sim - 10^{-11}i$.}.

\section{Conclusions}

We have examined the spectral (in)stability of Schwarzschild, RN, SdS and RNdS BH scalar QNMs under the influence of a P\"oschl-Teller bump that is added to the effective potential of the aforesaid metrics. This investigation is akin to environmental effects and their importance in BH spectroscopy, such as  simplistic models of accretion disks or matter shells in the exterior of the BH. Due to the variety of the spacetimes analyzed, we find numerous interesting results regarding the spectral stability of all families of QNMs. 

All static and spherically-symmetric BHs considered in this work possess a PS where null geodesics are trapped in unstable circular orbits. The excitation of the PS leads to the well-known QNMs. We find that the PS family of QNMs is spectrally unstable under the addition of the bump in the effective potentials of all the aforementioned BHs. Specifically, the further we place the bump, which is six orders of magnitude smaller than the PS peak, the more the PS QNMs are destabilized and disproportionately migrate in the complex plane towards perturbed QNMs with smaller oscillation frequencies and longer-lived timescales. This is a physical outcome of stable trapping of waves between the PS and the bump. The larger the distance is, the more time it takes for these modes to dissipate at the boundaries while inducing successively damped excitations of the PS, know as echoes~\cite{Cardoso:2016rao,Cardoso:2019rvt,Maggio:2020jml,Maggio:2021ans,Vlachos:2021weq,Chatzifotis:2021pak}. We further find that, for asymptotically-flat BHs, the spectral instability of PS QNMs leads to an effect akin to the anomalous decay of QNMs found in~\cite{Lagos:2020oek}, where for large enough bump positioning, the perturbed $\ell=0$ QNM becomes the dominant one, instead of the $\ell\rightarrow\infty$. Asymptotically dS BHs on the other hand do not exhibit the anomalous decay phenomenon, for the cosmological constants chosen here, due to the existence of the cosmological horizon which puts a limit the position of the bump and leads to an insufficient migration of QNMs. Nevertheless, PS QNM in SdS and RNdS BHs still are spectrally unstable since their migration is at least two to three orders of magnitude larger than the scale of the perturbing bump.

Besides the standard PS complex QNMs, BHs with a positive cosmological constant possess another family of modes that are purely imaginary. Those can even be found for pure dS spacetime~\cite{Lopez-Ortega:2006aal}. We have demonstrated that the dS family of modes is spectrally stable against the bump addition to the effective potential, as long as the bump is not placed too close to the cosmological horizon of the dS BH. Therefore, for physical choices of bump distance from the PS, we observe spectral stability, since we do not interfere with the asymptotic form of the effective potential. On the other hand, when we set the bump unphysically close to the cosmological horizon then the exponential decay of the effective potential is superseded by the bump's decay and this leads to a minimal spectral instability since we essentially introduce a perturbation of the scale of the observable Universe, which changes the boundary conditions. Thus, we conclude that dS modes are also spectrally stable for reasonable choices of bump locations.

The final family found for charged BHs, with or without a cosmological constant, is the NE family. These modes are purely imaginary for neutral scalar fields, are highly subdominant when the BH charge is not near extremality, and appear in the spectrum when the Cauchy and event horizon tend to coincide. At these near-extremally-charged BHs, the NE family becomes the dominant one that decays last in the ringdown waveform. We find that the NE modes have a superior spectral stability than that found for dS modes, in the sense that no matter the bump distance, they are always perfectly stable. Therefore, we may conclude that imaginary modes are spectrally stable, though this may change if different kind of perturbations to the effective BH potential are considered (e.g., perturbations which are not position-dependent).

Our extensive analysis of all QNM families and their stability under bump perturbations can be further applied to a fundamental aspect of GR, namely the SCC conjecture. Since it has been found that SCC is violated for near-extremely-charged RNdS BHs when linear neutral and massless scalar fields are considered~\cite{Cardoso:2017soq}, we have examined if the addition of the bump and the corresponding spectral instability of PS QNMs may remove or at least reduce the parameter space of SCC violation. Unfortunately, the results regarding the SCC violation remain utterly unchanged when realistic environmental effects are taken into account, due to the spectral stability of dS modes for small cosmological constants, and NE modes, as well as an inefficient migration of the PS QNMs for large cosmological constants where they dominate the ringdown signal. Nevertheless, we cannot exclude that the PS mode will eventually migrate and become dominant when the cosmological constant becomes arbitrarily small, though in this case the dS fundamental mode becomes even more long-lived. At best, SCC might be respected for extremely small cosmological constants, though this regime of the parameter space is practically impossible to probe numerically.

Regardless, in a recent study concerning the pseudospectrum of SdS BHs and the spectral instability occurring in both SdS and RNdS geometries under deterministic perturbations to the effective potential~\cite{Sarkar:2023rhp}, it has been found that no matter if the dS or NE modes are dominant, spectral instabilities affect these families, in contrast to what we find. This might occur due to the nature of the perturbation used in~\cite{Sarkar:2023rhp} that is position-independent, contrary to ours that is position-dependent. Furthermore,~\cite{Sarkar:2023rhp} used larger perturbations compared to ours, and the migration of the QNM spectrum seems compatible with the perturbation amplitude. Ultimately, the stability of dS and NE modes, when dominant, has to be analyzed with the pseudospectrum, that is oblivious to the form of the perturbation introduced to the operator that describes the QNM system, in order to unequivocally understand the spectral-stability properties of purely imaginary dominant families.

We note that even though the SCC conjecture is violated at the linear level in RNdS, nonlinear~\cite{Luna:2019olw}, quantum mechanical~\cite{Hollands:2019whz} and other mathematical treatments of the problem~\cite{Dafermos:2018tha} have shown that SCC can be healed, therefore the linear analog of the SCC is not enough to decide the ultimate fate of Cauchy horizons and the deterministic nature of GR. Finally, it is worth noticing that for astrophysically-relevant vacuum Kerr-dS BHs the SCC conjecture is respected for scalar and gravitational perturbations \cite{Dias:2018ynt}, due to the complex QNM family of Kerr-dS BHs (analogous to the PS family in spherically-symmetric spacetimes considered here). As we observed here, for scalar (as well as for gravitational~ \cite{Cheung:2021bol}) perturbations this family is the one that migrates to smaller (in absolute value) imaginary parts. Wee could therefore extrapolate that the SCC in Kerr-dS will continue being respected since the formation of quasibound states, due to the minuscule environmental bump, always tends to become longer-lived and $\beta$ will become even smaller than $1/2$. Even so, a proper numerical QNM analysis for Kerr-dS has to be performed in order to check our conjecture.

\begin{acknowledgements}
The authors would like to warmly thank Sumanta Chakraborty for helpful comments.
A.C. acknowledges hospitality from the Gravity group of the Physics department at Sapienza Università di Roma.
K.D. and P.P. acknowledge financial support provided under the European
Union's H2020 ERC, Starting Grant agreement no.~DarkGRA--757480, under
MIUR PRIN (Grant 2020KR4KN2 “String Theory as a bridge between Gauge Theories and Quantum Gravity”) and FARE (GW-NEXT, CUP: B84I20000100001, 2020KR4KN2) programmes, and support from the Amaldi Research Center funded by the MIUR program ``Dipartimento di Eccellenza" (CUP:~B81I18001170001). This work was supported by the EU Horizon 2020 Research and Innovation Programme under the Marie Sklodowska-Curie Grant Agreement No. 101007855.
\end{acknowledgements}

\appendix

\section{Bumps versus wells}\label{AppA}

\begin{figure*}[t]
	\centering
	\includegraphics[scale=0.375]{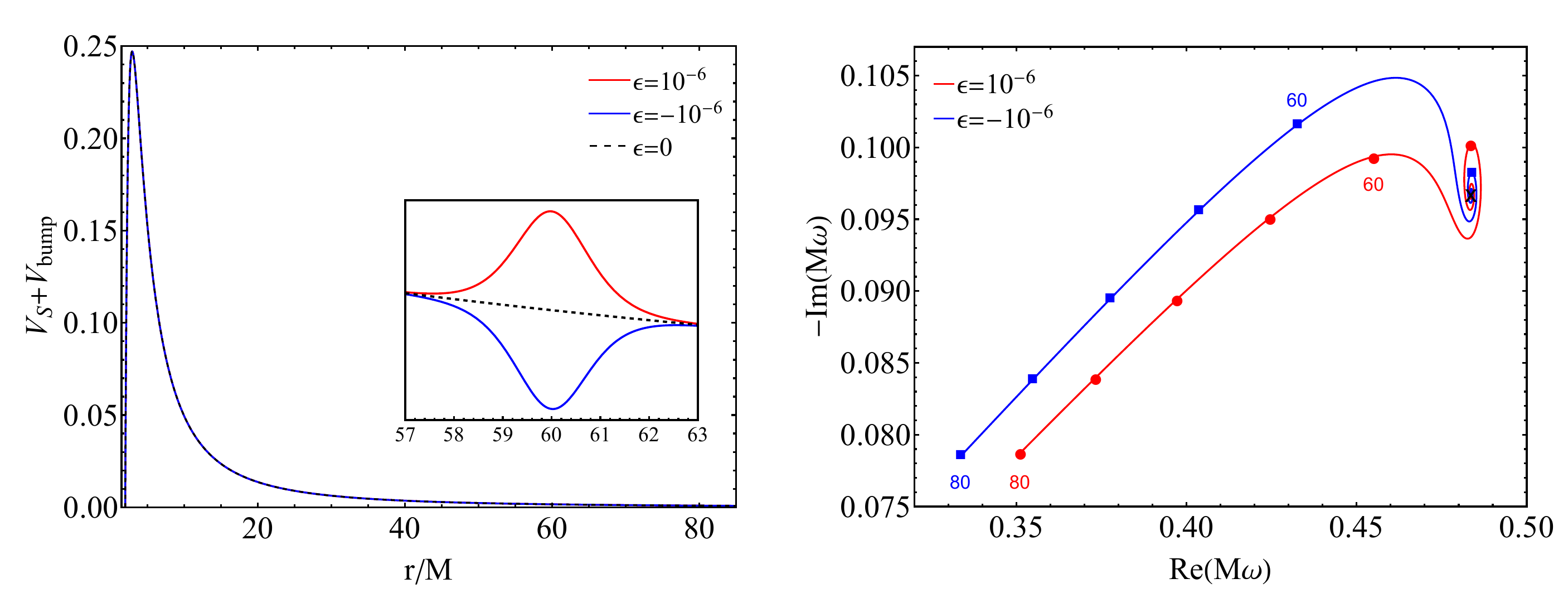}
	\caption{Left: Effective potential~\eqref{perturbed_potential} for a Schwarzschild BH perturbed by a bump ($\epsilon=10^{-6}$, red curve) and a well ($\epsilon=-10^{-6}$, blue curve). The dashed black curve corresponds to the unperturbed effective potential of scalar fields in Schwarzschild spacetime. Right: Migration of $\ell=2$ PS QNMs of a perturbed scalar field potential with a bump ($\epsilon=10^{-6}$, red curve) and a well ($\epsilon=-10^{-6}$, blue curve). In both cases, the red circles (blue squares) represent different values of $a/M$ for the cases of both positive and negative $\epsilon$. The black cross represents the fundamental unperturbed $\ell=2$ PS QNM.}\label{positive_vs_negative_bump}
\end{figure*}

A spherical matter shell surrounding a BH has the potential to give rise to a bump, similar to those we have considered in this work. If we consider a matter shell with a P\"oschl-Teller-like energy density (see Ref.~\cite{Cheung:2021bol}), we can integrate it to find the mass within the shell.

When the newly found (analytic or numerical) spacetime, in the form found in Ref.~\cite{Cardoso:2021wlq}, is perturbed by linear scalar fields, the matter shell will modify the effective potential as follows~\cite{Barausse:2014tra,Cheung:2021bol,Cardoso:2021wlq}:

\begin{equation}\label{potential_matter}
    V(r)=f(r)\left(\frac{\ell(\ell+1)}{r^2}+\frac{2m(r)}{r^3}-\frac{m^\prime(r)}{r^2}
    \right),
\end{equation}
where $m(r)$ is the radially-dependent mass of the shell and $f(r)$ is the respective lapse function of spacetime that includes the backreaction of matter shell as in~\cite{Cardoso:2021wlq}. Interestingly enough, the potential~\eqref{potential_matter} does not lead to a positive bump at the position of the matter shell, at least as in the case of a Schwarzschild BH as found in~\cite{Cheung:2021bol}, where axial gravitational perturbations were considered, but rather, the resulting potential forms a negative well at the would-be bump location. This may lead to a completely different strength of QNM migration with respect to $a/M$, or no migration at all for PS modes, and jeopardize the main results of our study.

In this appendix, we show that the sign of $\epsilon$ in Eq.~\eqref{bump} does not play a role in the migration of PS modes. We demonstrate in Fig.~\ref{positive_vs_negative_bump} that, no matter if the matter shell introduces a bump or a well, that can in principle be introduced by a realistic methodology to construct a spacetime with the properties of the studied effective potentials in the main text, the qualitative behavior of PS QNMs is strikingly equivalent. Therefore, we rest assured that the results presented in the main text regarding spectral instabilities and the SCC are valid.

\section{Cauchy horizon stability and the modern formulation of the SCC}\label{beta_proof}

Even though exact neutral and charged BH solutions surrounded by matter shells and halos have been recently constructed analytically for asymptotically flat spacetimes~\cite{Cardoso:2021wlq,Cardoso:2022whc,Feng:2022evy}, let us consider that such solutions exist for asymptotically dS geometries as well. The regularity condition for extensions beyond the Cauchy horizon is found by considering a backreacting matter field that describes the matter shell as a spherically-symmetric general fluid, with energy density $\rho$ and pressure $p$, and another one that describes the scalar field $\Psi$ considered as a probe to the metric at the Cauchy horizon vicinity.

If we consider Einstein's equations with a cosmological constant
\begin{equation}\label{field equations}
    G_{\mu\nu}+\Lambda g_{\mu\nu}=8\pi(T^\text{scalar}_{\mu\nu}+T^\text{shell}_{\mu\nu}),
\end{equation}
where schematically, $G_{\mu\nu}\sim\Gamma^2+\partial\Gamma$ with $\Gamma$ the Christoffel symbols that are proportional to $\partial g_{\mu\nu}$ (we have dropped the indices for simplicity), $T^\text{scalar}_{\mu\nu}\sim(\partial\Psi)^2$ is the scalar field's energy momentum tensor and $T^\text{shell}_{\mu\nu}\sim\rho u^2+\partial\rho/\partial g^{\mu\nu}-\rho g_{\mu\nu}$ is the energy-momentum tensor of a generic spherically-symmetric matter shell with conserved four-velocity $u^\alpha u_\alpha=u^2$ equal to $1$ (or $-1$ according to the signature convention)~\cite{Schutz:1970,Alho:2023ris}.

If we assume that $g_{\mu\nu}$, $\Psi$ are not necessarily twice-differentiable at the Cauchy horizon (the matter shell that depends on $\rho$ is static, spherically-symmetric and is only allowed to exist outside the BH so it does not contribute to the regularity condition at the Cauchy horizon), we can still make sense of Eq.~\eqref{field equations} by multiplying it with a compact support, smooth, function $\psi_0$ (infinitely-differentiable) and integrate it over the volume $\mathcal{V}$ at the Cauchy horizon, such that:
\begin{align}\nonumber
    &\int_\mathcal{V} d^4x\sqrt{-g}\left(\Gamma^2+\partial\Gamma\right)\psi_0+\int_\mathcal{V} d^4x\sqrt{-g}\Lambda g_{\mu\nu}\psi_0=\\\label{weak solution}
    &8\pi\left(\int_\mathcal{V} d^4x\sqrt{-g}(\partial\Psi)^2\psi_0+\int_\mathcal{V} d^4x\left(\rho+\rho^\prime-\rho g_{\mu\nu}\right)\psi_0\right).
\end{align}
The purpose of the test function $\psi_0$ is to carry away derivatives from fields without leaving non-zero terms when performing integrations by parts. Since $\psi_0$ is smooth, it will always be integrable no mater how many derivatives we carry out to it. If Eq.~\eqref{weak solution} is satisfied for any smooth function $\psi_0$, then we have a weak solution of Eq.~\eqref{field equations}.

For the left hand side of the weak field equations to be integrable, the square-integrability of the Christoffel symbols, or equivalently the integrability of the derivative of the metric tensor squared $\int_{\mathcal{V}}\sqrt{-g}(\partial g_{\mu\nu})^2 \psi_0 d^4x<\infty$, needs to be satisfied\footnote{Note that the term in Eq.~\eqref{weak solution} $\int_\mathcal{V}d^4x\sqrt{-g}\partial\Gamma\psi_0\sim\int_\mathcal{V}d^4x\sqrt{-g}\Gamma\partial\psi_0$ is integrable at the Cauchy horizon, where $\Gamma\sim\partial g_{\mu\nu}$, since the derivative of the Christoffel symbols can be transferred to $\psi_0$, thus the term that defines the regularity at the Cauchy horizon is $\Gamma^2\sim(\partial g_{\mu\nu})^2$.}. In a similar manner, the right hand side of~\eqref{weak solution}, includes both the linear scalar field and the matter shell energy-momentum tensors. For the scalar field, the requirement for square-integrability of derivatives of it guarantees that the part of its energy-momentum tensor leads to sufficient regularity at the Cauchy horizon when $\int_{\mathcal{V}}\sqrt{-g}(\partial \Psi)^2 \psi_0 d^4x<\infty$. The energy-momentum tensor of a general spherically-symmetric fluid that describes the matter shell (the last term on the right hand side of Eq.~\eqref{weak solution}) has no contribution to the Cauchy horizon's regularity since the energy density $\rho$ is associated with the energy/mass stored in the shell and not with that of an infalling observer. Thus, this terms is clearly integrable and does not contribute further to the regularity requirements for weak solutions at the Cauchy horizon.

At the Cauchy horizon, the scalar wave equation (in outgoing Eddington-Finkelstein coordinates $u\equiv t-r_*$ that are regular there) obtains two asymptotic solutions, namely $\Psi^{(1)}\sim e^{-i\omega u}$ (which is regular at the Cauchy horizon) and $\Psi^{(2)}\sim \Psi^{(1)}e^{-2i\omega r_*}$. Since the tortoise coordinate at the Cauchy horizon is approximately
\begin{equation}
    r_*=\int f^{-1}(r) dr\sim \frac{\log|r-r_-|}{f^\prime(r_-)},
\end{equation}
where $f^\prime(r_-)<0$, the potential non-smoothness at the Cauchy horizon arises from the solution $\Psi^{(2)}$, where
\begin{align}\nonumber
    \Psi^{(2)}&\sim e^{-2i (\omega_R+i\omega_I) r_*}=e^{-2i (\omega_R+i\omega_I) \log|r-r_-|/f^\prime(r_-)}\\\nonumber
    &\sim |r-r_-|^{i (\omega_R+i\omega_I) /\kappa_-}\\\label{regularity}
    &\sim |r-r_-|^{i\omega_R/\kappa_-}|r-r_-|^{-\omega_I/\kappa_-},
\end{align}
where we have defined the QNMs as $\omega=\omega_R+i\omega_I$, with $\omega_R$ the real part and $\omega_I<0$ the factor of the imaginary part. The first factor in~\eqref{regularity} is purely oscillatory and does not contribute to the regularity of solutions, therefore the only factor that matter is $|r-r_-|^\beta$, where $\beta\equiv \text{inf}(-\text{Im}(\omega_{n\ell}))/\kappa_-$, and $\text{inf}(-\text{Im}(\omega_{n\ell}))$ is the lowest-lying QNM of all families that defines the late time behavior of the ringdown signal.

Since the requirement for regularity at the Cauchy horizon is solely defined by the integrability of $(\partial \Psi^{(2)})^2$, we obtain 
\begin{align}\nonumber
    \int_\mathcal{V}(\partial_r \Psi^{(2)})^2 dr&\sim \int_\mathcal{V} |r-r_-|^{2\beta-2} dr\\
    &\sim\frac{|r-r_-|^{2\beta-1}}{2\beta-1}<\infty.
    \label{beta2}
\end{align}
For Eq.~\eqref{beta2} to be satisfied at the Cauchy horizon we require that $\beta>1/2$. Thus, we anticipate that the regularity condition $\beta>1/2$ that allows for weak extensions beyond the Cauchy horizon to remain the same for spherically-symmetric constructions of BHs with matter shells.

\bibliography{biblio}

\end{document}